\documentclass[preprint,10pt,number]{elsarticle}
\usepackage[utf8]{inputenc}
\usepackage{amssymb}
\usepackage{amsthm}
\usepackage{amsmath}
\usepackage{adjustbox}
\usepackage[section]{placeins}
\usepackage{color}
\usepackage{wrapfig}
\usepackage{etoolbox,refcount}
\usepackage{multicol}
\usepackage{geometry}
\usepackage{grffile}
\usepackage{setspace}
\usepackage[final]{pdfpages}
\usepackage{sectsty}
\sectionfont{\fontsize{15}{20}\selectfont}
\subsectionfont{\fontsize{13}{15}\selectfont}
\subsubsectionfont{\fontsize{11}{15}\selectfont}
\usepackage{graphicx}
\usepackage{caption}
\usepackage{subcaption}
\usepackage[superscript]{cite}
\usepackage{xcolor}
\usepackage{datetime}
\usepackage{gensymb}
\usepackage{textcomp}
\usepackage[font=normalsize,labelfont=bf]{caption}

\definecolor{outputorange}{rgb}{0.95, 0.44, 0.16}
\definecolor{inputgreen}{rgb}{0.46, 0.56, 0.31}
\journal{Nature Communications}
\usepackage{fontspec}
\newfontfamily\corsiva{Monotype-Corsiva-Regular.ttf}

\begin{document}
\begin{frontmatter}

\title{\Large A physical adaptive material motor unit neural network: a hygromorph composite material machine}

\author{Charles de Kergariou$^a$*, David Correa$^b$, Adam Perriman$^c$$^d$, Helmut Hauser$^e$$^f$, Fabrizio Scarpa$^a$}
\address{$^a$Bristol Composites Institute, School of Civil, Aerospace and Mechanical Engineering, University of Bristol, University Walk, Bristol BS8 1TR, United Kingdom

$^b$School of Architecture, University of Waterloo, 7 Melville Street South, Cambridge, Ontario, N1S 2H4, Canada

$^c$Research School of Chemistry and John Curtin School of Medical Research, Australian National University, Canberra ACT2601, Australia

$^d$School of Cellular and Molecular Medicine, University of Bristol, University Walk, Bristol BS8 1TD, United Kingdom

$^e$School of Engineering Mathematics and Technology, University of Bristol, Bristol, United Kingdom 

$^f$Bristol Robotics Lab, Bristol, United Kingdom

*Corresponding author.
Email address: charles.dekergariou@bristol.ac.uk}

\begin{abstract}
{\large Advances in novel materials science enable structures to function as intelligent machines by embedding memory and learning capabilities directly into materials. Our work introduces a physical adaptive material motor unit neural network, leveraging a new generation of controllable actuators composed of wood- and carbon black-based composites, sensitive to temperature and relative humidity. These material actuators are assembled into a motor unit-like structure inspired by muscle contraction trigger, forming an intelligent machine capable of dynamic shading control that can be used, for example, in buildings. The machine is governed by a neural network trained on over 350 experimental data points collected under diverse environmental conditions. By establishing a new data-aware backpropagation training, we show that the machine predicts shading responses and learns to predict appropriate behaviour incrementally as the database expands. We also demonstrate the ability of the machine to optimise configurations to achieve similar shading outputs under two distinct conditions.}

\end{abstract}

\end{frontmatter}


\begin{figure}[h]
    \centering
    \includegraphics[width=0.983\textwidth]{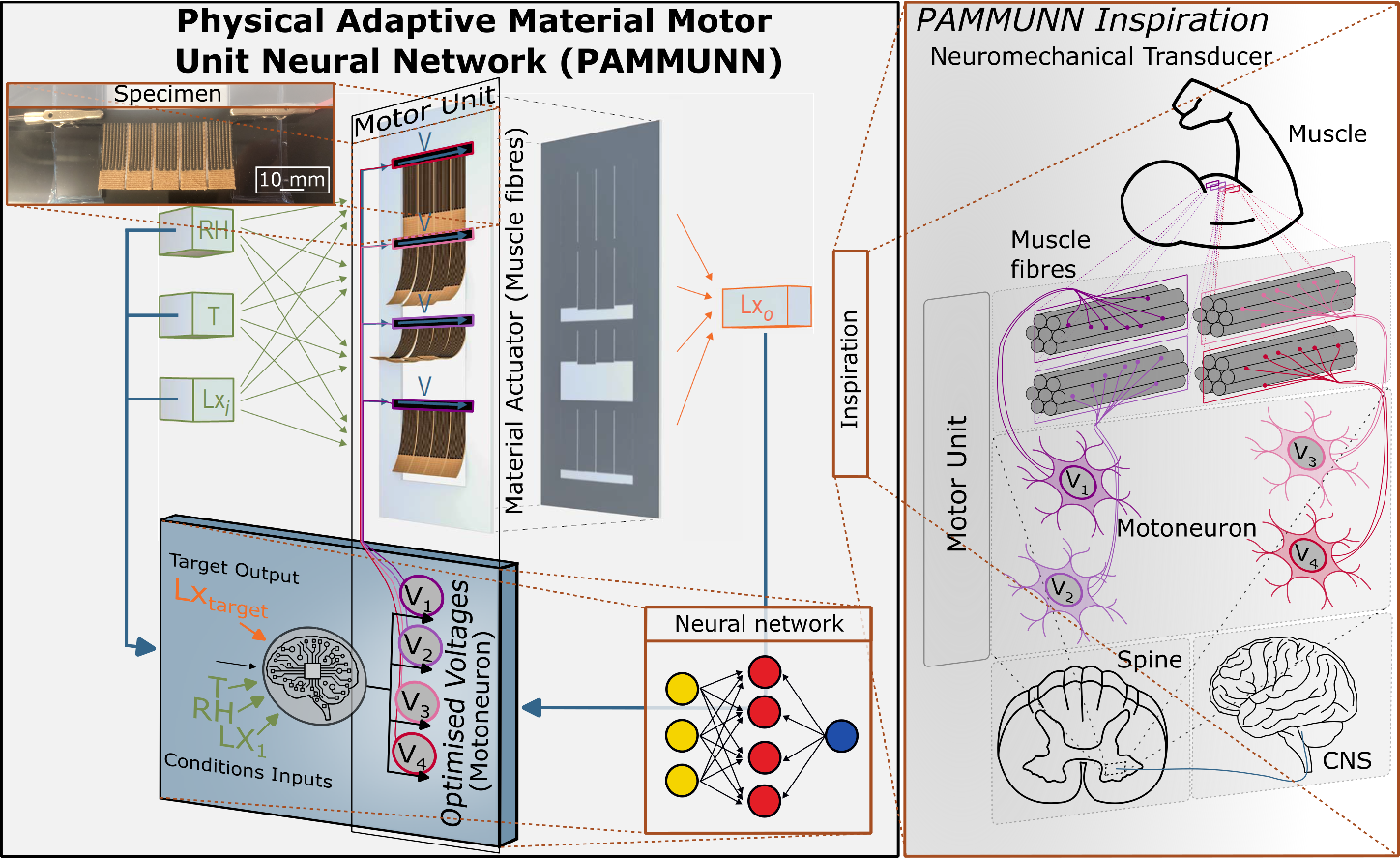}
    \vspace{-3mm}
    \caption*{\textbf{Graphical Abstract} }
    \label{fig: graphical abstract}
\end{figure}

\section{Table of Content}
This study introduces a new class of material-based intelligent machines. It presents a Physical Adaptive Material Motor Unit Neural Network, an assembly of smart actuators that enables machines to learn through environmental interaction. The material-based machine regulates light through a self-actuating facade, which opens in response to temperature and humidity.

\section{Introduction}
Material actuators are increasingly designed following  nature-inspired mechanisms \cite{Zhang2022UnperceivableCones,Chung2026RoboticElastomers} that enable the use of available energy in the environment to create motion \cite{Cheng2024Weather-responsive4D-printing}. The material within the actuator stores energy \cite{Aubin2022TowardsEnergy} and turns environmental stimulus, such as humidity change, into mechanical energy \cite{deKergariou2025Hygromnemics:PreConstraining}. Materials that are humidity-sensitive are called hygromorphs \cite{Cheng2024Weather-responsive4D-printing}. Other material actuators have been developed to achieve motion in response to environmental stimuli, such as change in temperature \cite{Hu1995SynthesisGels,deMarco20184DRobotics}, humidity \cite{deKergariou2025Hygromnemics:PreConstraining} or pH \cite{Zhou20244DControl}. Such energy transformation is used to create shading structures \cite{Cheng2024Weather-responsive4D-printing} that adapt to sun angles, smart clothes \cite{Biswas20214DChallenges}, smart biomedical implants \cite{Ding20254DApplications} or sensors that provide a kinetic response to environmental changes \cite{Teng20254DFruits}. However, three key aspects hinder the widespread adoption of technologies based on  material actuation, in particular for temperature/humidity triggers. The first challenge is the reduced scalability of the output of the actuation, such as amplitude \cite{Ding20254DApplications, Liu2025AdvancesApplications} and load carried \cite{Kergariou2025EffectiveActuators}. This is still a problem, despite the existence of large printing/additive manufacturing capabilities \cite{Zastrow20203DStronger}. The second problem is the lack of precise control of the output (e.g. actuating amplitude) for the wide variety of environmental conditions such actuator will be subjected to \cite{Reichert2015MeteorosensitiveResponsiveness}.  A third challenge is their partial irreversibility, which complicates precise actuation control \cite{Correa20204DMovement}. 

\begin{figure}[h]
    \centering
    \includegraphics[width=0.683\textwidth]{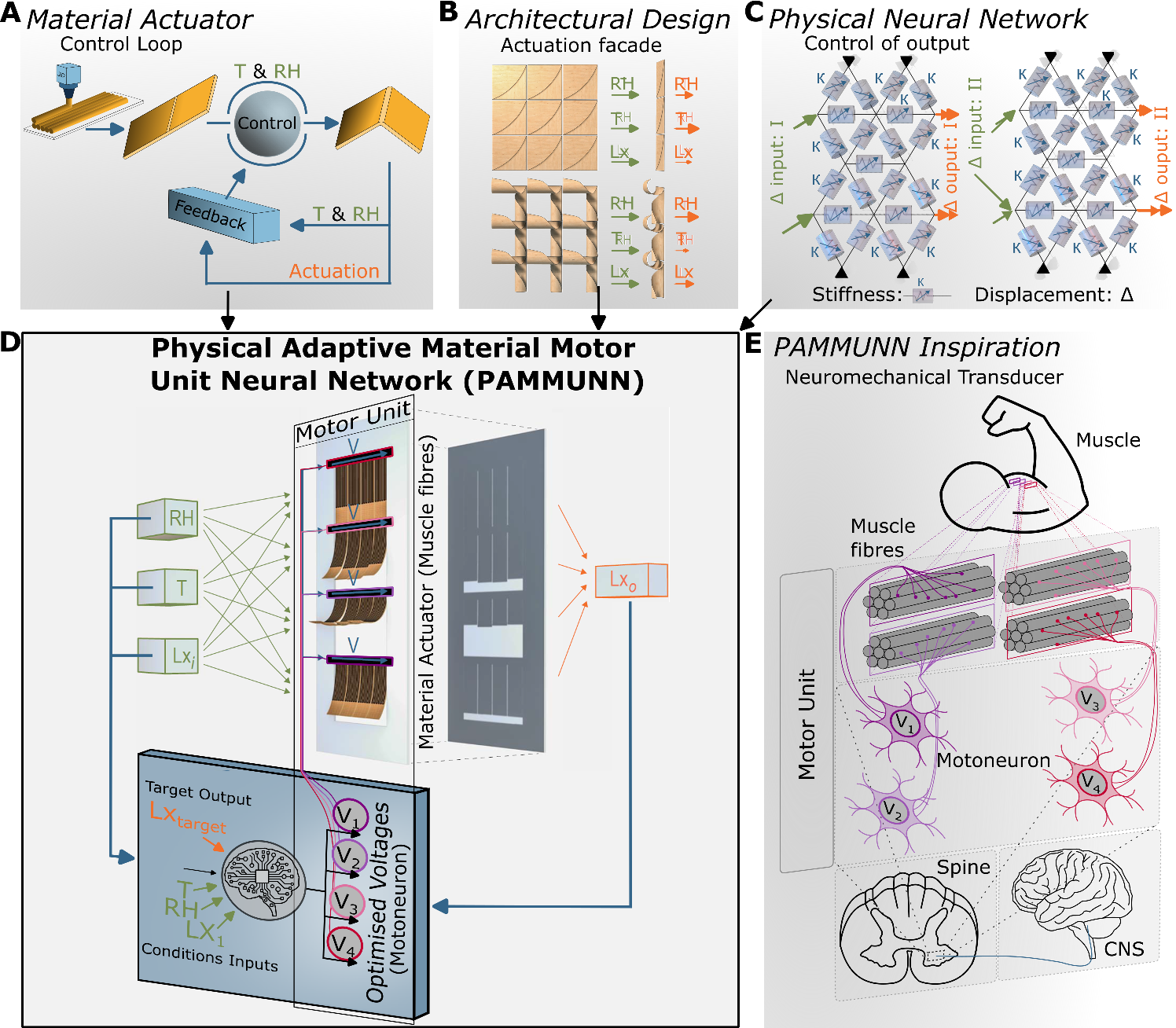}
    \vspace{-3mm}
    \caption{ \textbf{Concept and inspiration of the physical adaptive material motor unit neural network.} Three main types of technology are combined here to conceptualise PAMNN:\textbf{A,} Material Actuation \cite{deKergariou2025Hygromnemics:PreConstraining,MoralesFerrer2024MultiscaleStructures} and its control \cite{Hwang2025Closed-LoopExecution,Ji2024Closed-loopRobots}, \textbf{B,} a self-shading Architectural Design \cite{Cheng2024Weather-responsive4D-printing,Vazquez2024AEnvelopes} and \textbf{C,} a Physical Neural Network \cite{Momeni2025TrainingNetworks,Spall2022HybridNetworks} with the schematic of a Mechanical Neural Network by Lee et al. \cite{Lee2022MechanicalBehaviors}. \textbf{D,} the PAMNN-inspired structure of this study consists of four humidity-triggered actuators, whose deformation is controlled by heating a conductive composite. The structure imitates a one layer NN which turns three inputs (T: Temperature, RH: Relative Humidity, Lx$_i$: Input light) into an output (Lx$_o$: Output light), which is represented by the light inside the box covered by the actuators. Due to its NN-based digital training, the output of the structure is predicted in a similar way to a feed-forward neural network, with voltage serving as control of the opening of the neurones. The structure retains actuation data, enabling it to continuously learn from the interactions between inputs and outputs. \textbf{E,} Inspiration for PAMMUNN in which the Central Nervous System (CNS) creates the inputs to control the force to generate movement \cite{Heckman2012MotorUnit} via the rate of actuation potential firing rate in the neurone \cite{Semmler2002MotorPerformance}. The information is transmitted through the spine to the motor units which is a neuromechanical transducer made of a motor neuron and the muscle fibres it triggers for contraction. }
    \label{fig: concept and inspiration}
\end{figure}

One way of solving the limited size of actuators is to split them in small sub-units assembled in arrays, whose output is the cumulative effect of all individual actuators. For instance, that is the concept behind self-shading facades \cite{Cheng2024Weather-responsive4D-printing,Yi2021PrototypingPanel}. However, programming an adaptive facade with  material actuators in a real building involves the interaction of a virtually infinite number of micro-climatic environments. This involves differences in sun exposure, height (i.e. air more humid near ground than at the top), temperature, proximity to vegetation, partial shading from other buildings, and interaction between actuators. A top-down approach to programming such system is practically not feasible, due to significant computational costs involved in predicting all possible combinations of conditions for each local situation. To account for the complexity of the environment, the  material actuator itself should \textit{remember} and \textit{process} information to adapt to the local conditions and give the required output. Hence, one layer of such actuators has the potential of behaving similar to a layer of a neural network, with each actuator being a node. Structures with multiple small actuators provide an opportunity of fine control of the output. Fig. \ref{fig: concept and inspiration} presents the sources of inspirations for the proposed artificial neural network machine made with material actuators, which we call a Physical Adaptive Material Motor Unit Neural Network (PAMMUNN).

Proposed concepts of Physical Neural Networks \cite{Momeni2025TrainingNetworks,Xu2026PhysicalTraining, Xue2024FullyNetworks, Lee2022MechanicalBehaviors} possess architectures, in which connections between artificial neurons are stiffness-based and make use of concentrated parameter elements \cite{Mei2024MechanicalNeurons}. Physical neural networks also feature topologies made of assemblies of connecting nodes, each of them autonomous from an energy perspective. The use of materials with intrinsic actuation capabilities is a first step to provide true autonomy. This aligns with the principles of \textit{physical control} \cite{Milana2025PhysicalRobotics} and \textit{morphological computation} \cite{Hauser2023LeveragingRobots}, leveraging oscillations, sequences and reactions for self-regulation. For instance, implemented with compact, autonomous robotic systems \cite{Aubin2022TowardsEnergy,Wehner2016AnRobots}, and also actuating facades and architectural designs \cite{Takahashi2024StructuralHygroShell, Alhafnawi2026ArchitecturalExpression}. Physical Adaptative Material Neural Networks embody three key design principles: i) be made of identical  material actuating neuronal nodes; ii) be able to find optimal actuating configurations to reach targeted outputs from given inputs; and iii) demonstrate learning at the material actuation level to boost reversibility and adaptability. The ultimate goal of the development of physical neural networks is to create a  machine with no traditional computing hardware (e.g. GPUs) to achieve the three design principles mentioned above \cite{Momeni2025TrainingNetworks}.   

The motor neuron unit (PAMMUNN) we propose in this work contributes towards the development of physical neural networks with adaptation, energy autonomy and self-regulation at the intrinsic material level.  As described in Fig. \ref{fig: concept and inspiration}D and E, the PAMMUNN is inspired by a biological motor unit (combination of a motoneuron and the muscle fibres whose contraction it controls). Similarly to some biological systems, the PAMMUNN presented in this study is capable of receiving environmental inputs to control shading output. To attain this performance, the design of this PAMMUNN is based on the use of natural fibres reinforced polymer composites, which are responsive to humidity \cite{Kergariou2025EffectiveActuators}. As a consequence, we define a new humidity responsive (hygromorph) composite architecture based on combinations of wood/thermoplastic and carbon black/PLA filaments deposited using filament fused fabrication. The wood fibre-dominated actuation \cite{deKergariou2023TheChallenges, deKergariou2025Hygromnemics:Pre-Constraining} of this adaptive material is controlled via electrical power heating \cite{MoralesFerrer2024MultiscaleStructures,LaleganiDezaki2023MagnetorheologicalActuators}. With this material, we create an assembly of four identical actuating blocks which are equivalent to the nodes of a neural network capable of controlling the penetrating light (Figure \ref{fig: concept and inspiration}D).  Physical training in biological motor units adapts the fire rate of the actuation potential and synchronisation of the motor units \cite{DelVecchio2019TheCoding,Semmler2002MotorPerformance}. Like in motor units, the PAMMUNN has its mechanical actuation controlled via electric voltage by a centralised neural network \cite{Heckman2012MotorUnit}. We also employ physical training, allowing the PAMMUNN structure to learn adaptive behaviors through iterative physical updates. To do so, a physics-aware back-propagation-inspired training is developed by linking inputs (temperature, humidity, environmental light) to the output light by controlling the heating of the  material actuators.  comfort\cite{Takhmasib2023Machine-learnedComfort,Alhafnawi2026ArchitecturalExpression}. The PAMMUNN learns with computation, then transmits the new knowledge to the physical system. We also demonstrate that this PAMMUNN biocomposite hygromorph composite machine learns its behaviour by acquiring data.

\section{Results}

\subsection{PAMMUNN using Data-Aware Back-Propagation Training}
\subsubsection{Database: acquisition}
\begin{figure}[h]
    \centering
    \includegraphics[width=0.8683\textwidth]{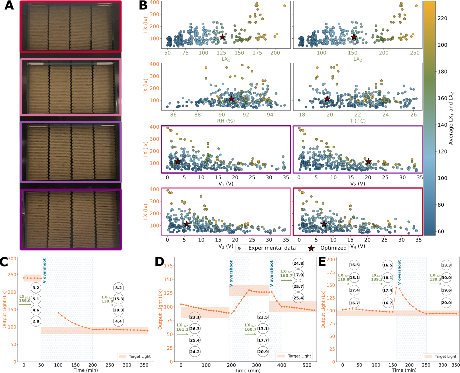}
    \vspace{-3.5mm}
    \caption{\textbf{Example of databases an output light evolutions for scenarios 1 and 2.} \textbf{A}, Image of the actuating structure for the 4*1 down actuation setup. \textbf{B}, Database created from testing the QSDF aperture set up from Fig. \ref{fig: Stacking Sequence}. Black stars are examples of optimised input voltage via the Neural Network control system presented in Fig. \ref{fig: PAMMUNN via Physics-Aware Back-Propagation Training: Scenario 1 - Example 3} of the Supplementary Materials. \textbf{C}, Scenario 1: Example of one stage convergence obtained for the 4*1 down actuation setup. The blue circle filled section corresponds to the time in which the optimised voltages are artificially increased to reach the predicted output value faster. The orange bands show the range for the target output light intensity. The black circles present values of the voltage induced in each individual actuating specimen. The top value is related to the top actuating specimen presented in Fig. \ref{fig: Stacking Sequence}B and to the V1 values of Fig. \ref{fig:NN-like structure schematic}. The bottom value is related to the bottom specimen shown in Fig. \ref{fig: Stacking Sequence}B and the V4 value of Fig. \ref{fig:NN-like structure schematic}. \textbf{D} shows an example configuration for the scenario 1 with multi stages of optimisation. \textbf{E}, Example configuration for scenario 2.}
    \label{fig: database example}
\end{figure}

The first PAMMUNN prototype was created by assembling four novel hygromorphic actuators side by side to form a small-scale shading structure (see Methods section). This assembly of soft robotic actuator is called Quadruple Stacked Down Facing QSDF aperture setup (Fig. \ref{fig: Stacking Sequence}) and it was defined after initial study presented in Supplementary Materials. In the present paper \textit{structure} refers to the assembly of hygromorphic actuators and the DC power supply connected with them to control their actuation and \textit{machine} refers to the entire system with control. Fig. \ref{fig: database example}B presents an example of training database used for the model in Fig. \ref{fig:NN-like structure schematic}. Such database was used to train a data-aware back-propagation training to control the shading of PAMMUNN. We define in the method section data-aware back-propagation training as a new type of back propagation training method for physical neural network functioning in a similar manner to physics-aware back-propagation trainings \cite{Momeni2025TrainingNetworks} without the need for a digital model.

Fig. \ref{fig: database example}B shows the impact of different parameters on the output of the system (light going through). The input lights LX$_1$ and LX$_2$ show a strong correlation with the output light. In general, the larger the light input, the higher the light output. The temperature and humidity do not show a clear correlation with the output of the structure in Fig. \ref{fig: database example}B. Without pre-constraining, biocomposite-based actuators tend to bend with the increase of humidity \cite{Cheng2024Weather-responsive4D-printing,Reichert2015MeteorosensitiveResponsiveness}. While humidity influences actuation, voltage dominates this effect. The observed humidity range remains narrow (Maximum range between [85\%; 95\%] as seen in Fig. \ref{fig: database example}B) compared to the typical actuation ranges (Several tens of relative humidity percentages, the minimum range observed by a previous review \cite{deKergariou2023TheChallenges} is 40\%) for such soft robotic materials. The scatter of the data related to the four input voltages shows that the larger the voltage input, the lower the output light.

\subsubsection{PAMMUNN: Implementation}

We present scenarios to explore the ability of the optimisation technique and the performance of physical adaptive material neural network machine.

\textbf{Scenario 1}

In this scenario the light output is set to converge out of an interval of light into another one when the user decides to change the target desired light of the system. The optimised voltage values obtained from the data-aware back-propagation training drive the system to the target light interval, demonstrating how a self-shading structure dynamically adjusts its shading output based on operator input.
In the configuration presented in Fig. \ref{fig: database example}C, the input light is 139.0 (average between LX$_1$ and LX$_2$). The target output light (orange bands) after decision of the operator ranges from 80 Lx to 100 Lx. The voltage inputs initially are V$_1$ = 3.2 V, V$_2$ = 5.1 V, V$_3$ = 4.6 V and V$_4$ = 2.8 V (left circles in Fig. \ref{fig: database example}C). Due to this range of input voltage, the output light initially converges towards 239.4 lx (orange line between 0 and 50 minutes). Then, the neural network lowers the output light to 88.8 Lx, resulting from a new set of voltage values (V$_1$ = 8.2 V, V$_2$ = 15.3 V, V$_3$ = 29.3 V and V$_4$ = 4.4 V). The converged output light values show a 150.6 Lx gap—from the initial 239.4 Lx to the final 88.8 Lx in the targeted interval. This 150.6 Lx gap closed in under 150 minutes (blue circle region). The second configuration presented for this scenario (Fig. \ref{fig: database example}D) aims to show that the machine can adapt to smaller and multiple changes of targeted light outputs by the user. Fig. \ref{fig: database example}D demonstrates that the neural network-like physical machine reliably converges multiple times to the different target light output. Four additional example of this scenario are given in the Supplementary Materials (see Fig. \ref{fig: PAMMUNN via Physics-Aware Back-Propagation Training: Scenario 1 - Example 3}, \ref{fig: PAMMUNN via Physics-Aware Back-Propagation Training: Scenario 1 - Example 4}, and \ref{fig: PAMMUNN via Physics-Aware Back-Propagation Training: Scenario 1 - Example 5}).

\textbf{Scenario 2}

In this scenario, once convergence is achieved in the target light interval, the light input is altered—and the system re-converges to the original target interval between 80 Lx to 100 Lx (orange band in Fig. \ref{fig: database example}E). This scenario demonstrates how the system adapts autonomously when the input light changes—such as shifts in the sun’s position relative to the shading structure. Fig. \ref{fig: database example}E illustrates a scenario where, after converging to the 80–100 Lx target interval, the input light suddenly spikes—from 119.8 Lx to 139.3 Lx (average of LX1 and LX2). In this configuration, the QSDF aperture restores the output light to the target interval in under 70 minutes, achieving a 50 Lx reduction. Three other example of this scenario are presented in Supplementary Materials (see Fig. \ref{fig: PAMMUNN via Physics-Aware Back-Propagation Training: Scenario 2 - Example 2}, \ref{fig: PAMMUNN via Physics-Aware Back-Propagation Training: Scenario 2 - Example 3} and \ref{fig: PAMMUNN via Physics-Aware Back-Propagation Training: Scenario 2 - Example 4}).

\subsubsection{PAMMUNN: Incremental Learning}
\begin{figure}[h]
    \centering
    \includegraphics[width=0.9683\textwidth]{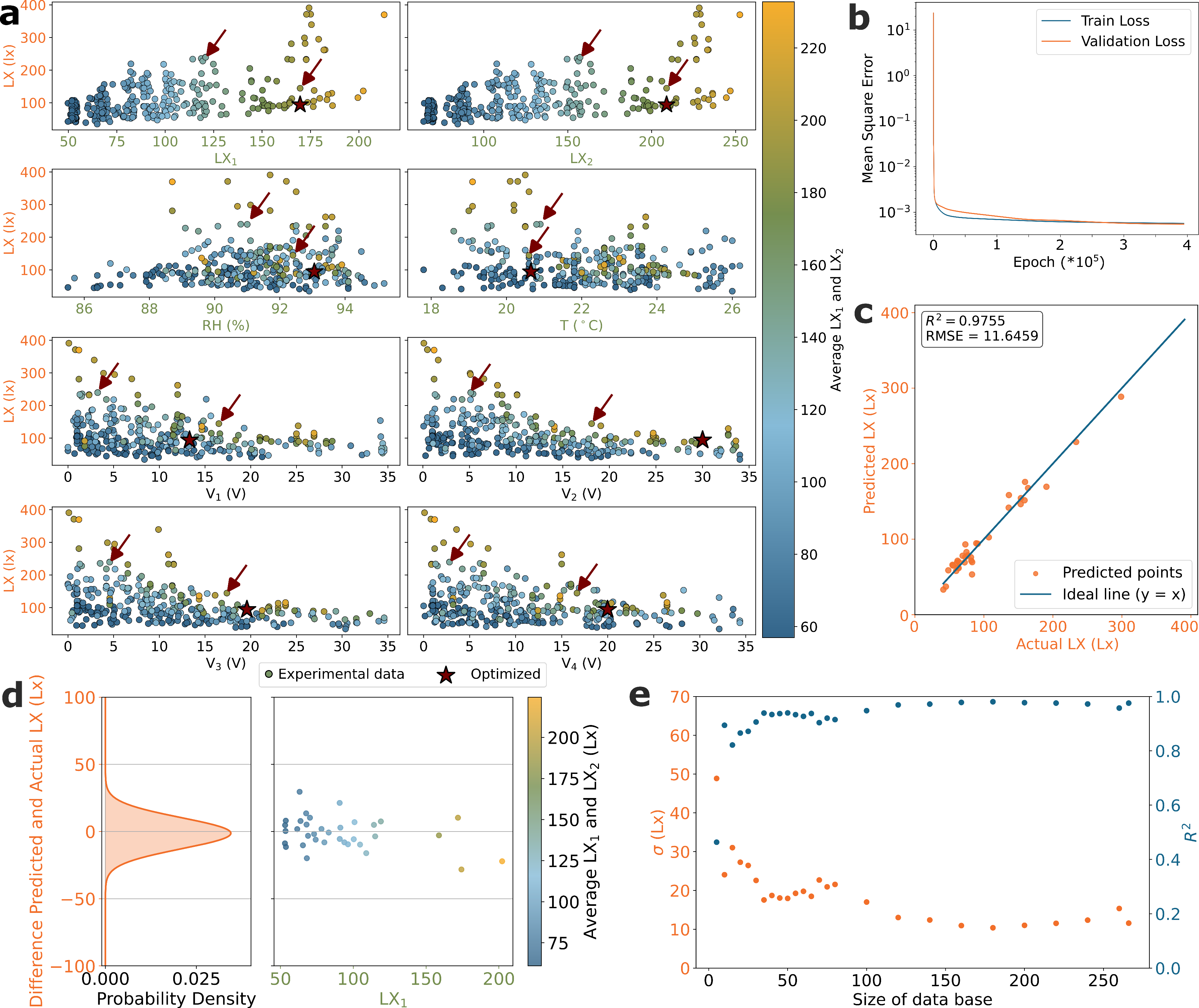}
    \vspace{-3mm}
    \caption{\textbf{Optimisation and results for incremental learning demonstration.} \textbf{A} Shows the database used to calculate the optimal voltages to reach the target interval. \textbf{B}, Evolution of the Error during training of the neural network with all the database considered. \textbf{C}, Actual versus Predicted data for the validation data used during training of the last iteration of the database. \textbf{D}, Gaussian distribution of the difference between the actual and predicted LX values for the validation data at the last iteration of the training. \textbf{E}, Evolution of the standard deviation of the difference between predicted and actual values and of the correlation coefficient  for every iteration of the training during demonstration of incremental learning for the machine created.}
    \label{fig: incremental training}
\end{figure}

In Fig. \ref{fig: incremental training}A, the downward red arrows indicate new data points added to the database, compared to the initial scenario in Fig. \ref{fig: database example}B. These arrows show the increase of database size for this new optimisation case potentially inducing an incremental learning for the machine. Fig. \ref{fig: incremental training} shows the different steps of the database training during demonstration of incremental learning. This test aims to present how an incremental increase in database size improves the quality of the prediction by the data-aware back-propagation training. The database increase incrementally in size as the machine operates it collected data of the relation between input and outputs parameters. Fig. \ref{fig: incremental training}B shows a sample convergence curve for the proposed model. 50 evaluation points were used in the data-aware back-propagation training. Fig. \ref{fig: incremental training}C compares the actual vs. predicted values, evaluating the performance of the trained neural network on these points of the database. From this data, the errors between actual and predicted values were calculated and plotted, with their Gaussian distribution shown in Fig. \ref{fig: incremental training}D. Two metrics were used to evaluate the machine’s adaptability, the correlation coefficient (R$^2$) and the standard deviation $\sigma$ of the related Gaussian distribution of the difference between actual and predicted values used in the neural network of the data-aware backpropagation training (Fig. \ref{fig: incremental training}C). The two metrics are plotted against the size of the database in Fig. \ref{fig: incremental training}E.

Fig. \ref{fig: incremental training}E demonstrates that the standard deviation of prediction errors drops sharply as the database grows to 120 points, then stabilises during the last validation to $\approx$10 Lx. In Fig. \ref{fig: incremental training}E, the correlation coefficient rises steeply up to 120 database points, then gradually converges to $\approx$0.98. These two evolutions highlight the learning ability of the machine. As the database increases in size the differences between the predicted and the actual values obtained reduces leading a better prediction ability of the machine. For instance, the Gaussian distribution of the difference between predicted and actual values shown in Fig. \ref{fig: incremental training}D reveals that four in five predicted values falls within a $\pm$14.9 Lx range of the target. This range reduces in size along the addition of tested data points proving learning ability of the machine.

\subsection{PAMMUNN: Physical-Mode}

Inspired by the neural network tests performed by Lee et al \cite{Lee2022MechanicalBehaviors}, Fig. \ref{fig:NN-like structure schematic}C and D present the same input under two different structural configurations. This physical-mode was implemented and tested with the conditions presented in Fig. \ref{fig: Example of NN-mode results}. Two other examples are presented in Supplementary Materials.

\begin{figure}[h]
    \centering
    \includegraphics[width=0.7658\textwidth]{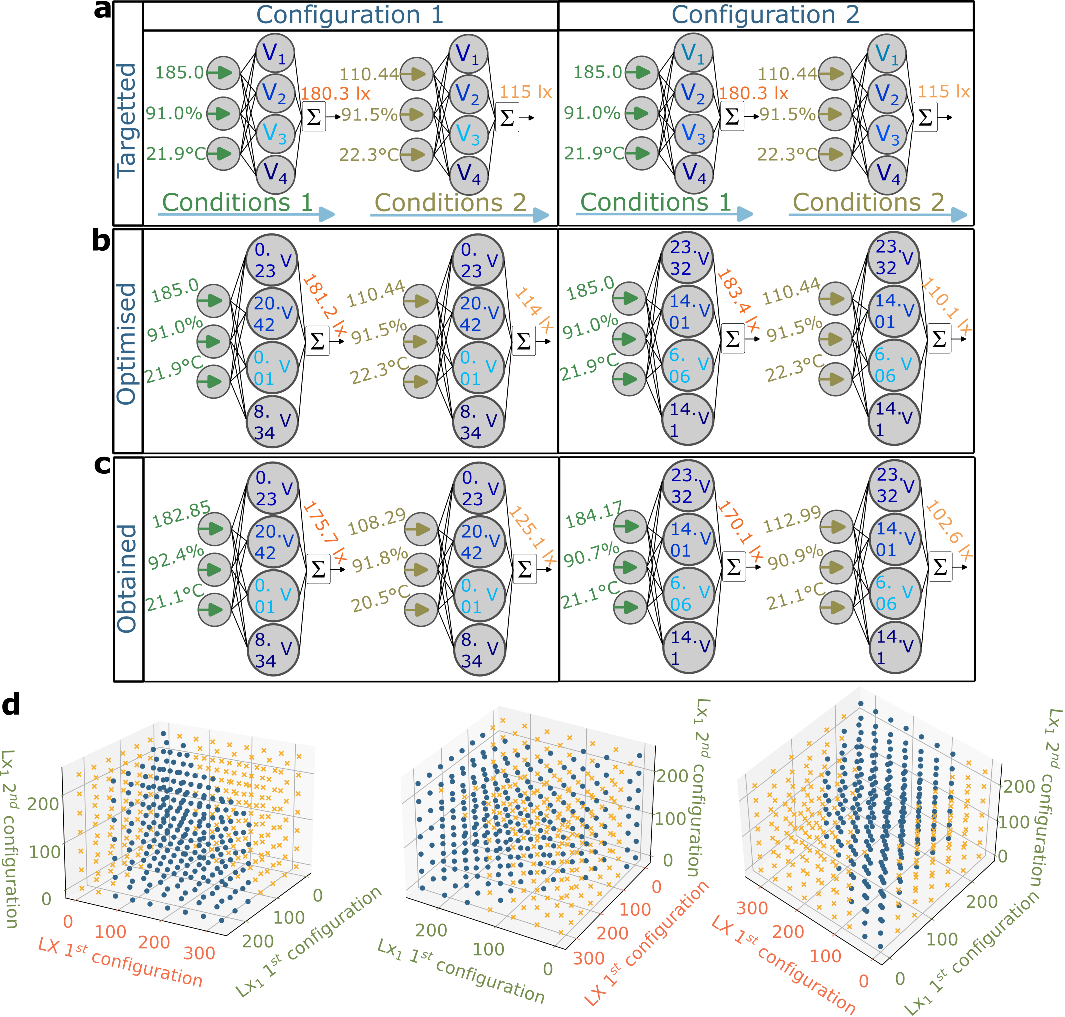}
    \vspace{-3mm}
    \caption{\textbf{Example of physical-mode operating.} \textbf{A,} Targeted output for given conditions. \textbf{B,} The optimized voltages are obtained from the specified conditions and targetted outputs. Due to the structure’s physical constraints, only certain output values are attainable. To this extend, the optimised output values differ from the targeted ones in \textbf{A,}. \textbf{C,} The optimised voltages previously calculated are implemented in the physical structure to obtain outputs close to the ones optimised. \textbf{D,} Design space for the physical-mode of the machine (three different angles of view). Blue dots corresponds to the configurations and conditions for which the shading structure developed found at least two sets voltages for the two targetted outputs given two inputs. This is a necessary condition to allow the setup to be used in this mode. The yellow crosses correspond to the configurations and targetted output for which the shading set up could not find two sets voltages allowing for the machine to function in physical-mode.}
    \label{fig: Example of NN-mode results}
\end{figure}

This figure shows that a difference exists between optimised output values in Fig. \ref{fig: Example of NN-mode results}B and those physically obtained in Fig. \ref{fig: Example of NN-mode results}C. These differences are 5.5 Lx, 11.1 Lx, 13.3 Lx and 7.5 Lx for the configuration 1 with conditions 1 and 2, and configuration 2 for condition 1 and 2, respectively. Two main reasons for these differences are the variabilities observed in the physical systems of the scenarios presented in the previous sections (see Fig. \ref{fig: database example} and the standard deviation introduced in Fig. \ref{fig: incremental training}). The second reason is the difference in terms of temperature, light and humidity observed between the two types of datasets, Optimised (Fig. \ref{fig: Example of NN-mode results}B) and Obtained (Fig. \ref{fig: Example of NN-mode results}C). For example, in configuration 2 under condition 1 (Fig. \ref{fig: Example of NN-mode results}C), both light and humidity fall below the optimised levels shown in Fig. \ref{fig: Example of NN-mode results}B. This artificially reduces the output, as reflected in the experimental data.

Fig. \ref{fig: Example of NN-mode results}D shows the design space of the physical machine behaving as in the physical-mode. Lower light input in one configuration limits the use of physical-mode, reducing viable options of sets of voltage to obtain the second configuration. This partly explains the difference of outputs between targetted values observed in Fig. \ref{fig: Example of NN-mode results}A and  the optimised one in Fig. \ref{fig: Example of NN-mode results}B. For instance,  in configuration 2 condition 1 the optimiser modified the output light from 180.3 lx to 183.4 lx to be able to fins a configuration able to come close to the required input and targetted outputs. However, in Fig. \ref{fig: Example of NN-mode results}D as the input light of the first configuration increase, proportion of second configuration achievable increases as well. This is explained by the fact that low sets of "low" voltages  will allows increasing number of large light input and, the sets of "high" voltages will permit low light outputs to be obtained. For similar reasons limited number of second configuration outputs are available when the output light of the first configuration is low. Low outputs required for the first configuration limits the possibility of using the physical configuration in physical-mode as it implies that one of the conditions will have low light inputs. It could also be due to the fact that at low output for the first configuration large voltages at all neurones limiting the adaptability of the soft robotic material for the second configuration. This variability is important in the context of this study as we proposed a 5 V mandatory difference between at least 5 of the voltages of the two configurations to display the physical-mode possibility of the machine. Expanding the actuating structure via addition of more opening enlarges the design space, enabling more aperture variations per input-output requirement.

\section{Conclusion}
This study investigates a physical actuator architecture made of adaptive biocomposite materials that approximates neural network behaviour. A new material actuator was designed using a wood-based thermoplastic and carbon black-reinforced PLA, and a novel geometry was defined for shading structures. The biocomposite actuator, with its fibre-dominated actuation structure, is activated by drying the fibres through heating from the conductive carbon black-reinforced PLA. Initially, two actuating configurations were compared, showing that downward-oriented specimens are preferable for creating self-shading structures than configurations with both upward and downward actuators. The structure designed for this study therefore includes four identical, parallel actuator apertures for shading. The shading output was generated by controlling voltage through the carbon-black reinforced PLA section, using environmental inputs like temperature, humidity and light. The light output of the adaptive material actuator was controlled using a newly defined \textit{Data-Aware Back-Propagation Training}, providing the first self-learning control system for material actuators, to the best of the authors' knowledge. A neural network trained with the new Data-Aware Back-Propagation provided the optimized control voltages. Neural network validation confirmed that four of five optimized voltages kept the light output within $\pm$14.90 Lx. The hygromorph machine's efficiency was tested in two scenarios, demonstrating the ability of the control technique to manage light output, whether the target changes or inputs vary. We were able to display incremental learning by increasing the size of the database as the machine operates.
\\

The Physical Adaptive Material Motor Unit Neural Network machine consists of actuating material neural nodes. It optimizes configurations from inputs and demonstrates learning digitally, not at the material level. The Data-Aware Back-Propagation Trained self-trained PAMMUNN also achieves two optimal configurations that convert two different inputs into similar outputs, proving ability to function as a neural network. In this mode, PAMMUNN reduces reliance on external digital and/or electronic systems, enabling future permanent learning based on the actuating material's physical properties.

\section{Methods}
\subsection{Printing and Prototype}
\begin{figure}[h]
    \centering
    \includegraphics[width=0.783\textwidth]{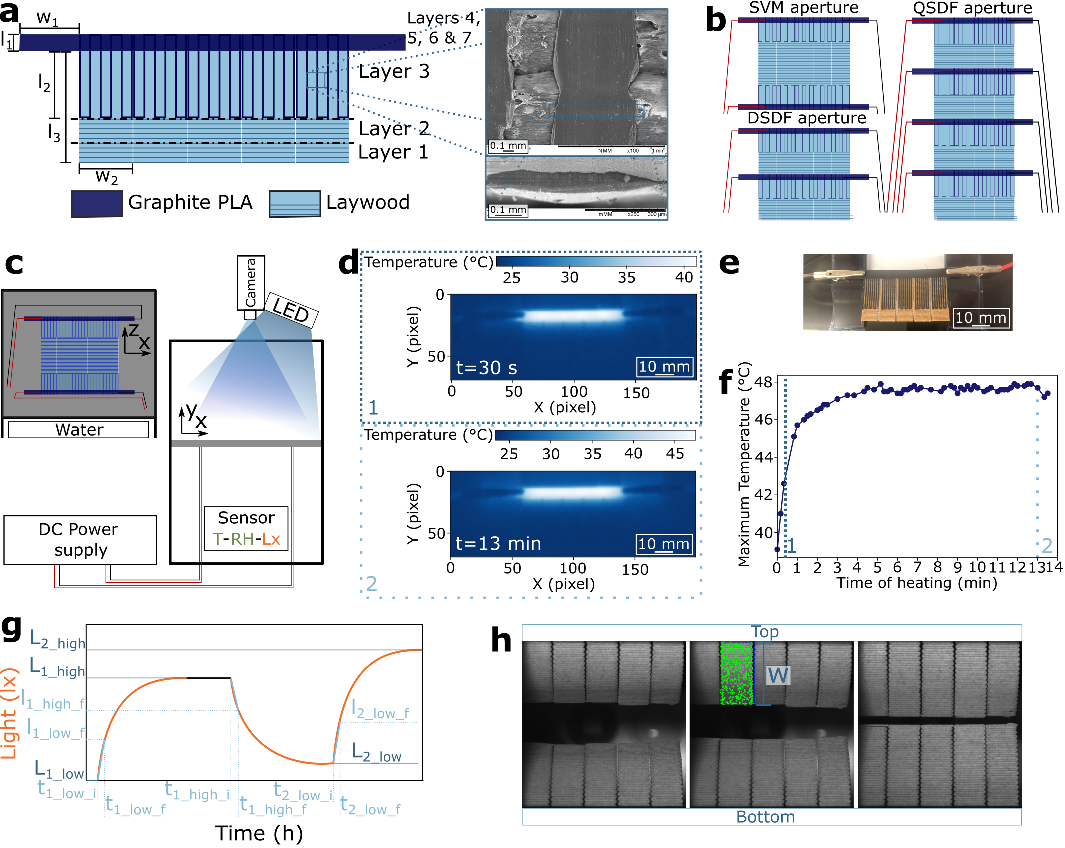}
    \vspace{-3mm}
    \caption{\textbf{Printed geometry, stacking sequence, SEM images, spatial architecture of the specimens fabricated and experimental setup implemented. Definition of speed reversibility and actuation strain.} \textbf{A}, Left schematic: Schematic of the stacking sequence of the specimen Right images: The top image is a top view SEM photo of the junction between the natural fibre and graphite filled PLA. \textbf{B}, Spatial architecture of the specimens for different tests named Symmetrical Vertically Mirrored (SVM) aperture, Double Stacked Down Facing (DSDF) aperture and Quadruple Stacked Down Facing (QSDF) aperture. The red and black lines on each side of the specimens are the electric connector to the DC power supply presented in Fig. \ref{fig: Stacking Sequence}C.
    \textbf{C}, Schematics of how the SVM aperture specimen of Fig. \ref{fig: Stacking Sequence}B is positioned. \textbf{D}, Example of thermal images recorded during the heating of one specimen. Top and Bottom images are recorded after 10 seconds and 13 minutes of heating, respectively. \textbf{E}, Image of the specimen heated by setting up the DC power supply with a 50.0 V input to record thermal images. \textbf{F}, Evolution of the maximum temperature with time when the specimen is heated with a 50.0 V input in the DC power supply.
    \textbf{G}, Definition of reversibility and speed parameters for soft robotic material actuation. \textbf{H}, Example of images recorded during experiment testing SVM aperture, after the voltage is set up to 50.0 V.}
    \label{fig: Stacking Sequence}
\end{figure}

The specimens are produced with Laywood meta 5 conductive carbon black-reinforced PLA (Kai Parthy and Protopasta) with the stacking presented in Fig. \ref{fig: Stacking Sequence}A. Layers 1 and 2 are made of unidirectional printing laywood. Layer 3 has also an unidirectional printing path but with a filling of 20\% and printed with Protopasta graphite reinforced PLA. Layers 4, 5, 6 and 7 are made with rectilinear printing paths of Protopasta carbon black PLA. The dimensions presented in Fig. \ref{fig: Stacking Sequence}A for the specimens are as follow: w$_1$: 9.5 mm, w$_2$: 9.5 mm, l$_1$: 5 mm, l$_2$: 22 mm and l$_3$: 30 mm. The composites are printed using a Prusa MK4 at a printing bed temperature of 80 $\degree$C and a nozzle temperature of 220 $\degree$C. The heights of the layers (0.4 mm between laywood layers and 0.2 mm between laywood and PLA layers) have been chosen to obtain a good contact between the two different soft robotic materials (Fig. \ref{fig: Stacking Sequence}A). The contact between the different layers of an actuators is key to control actuation and limit irreversibility \cite{Correa20204DMovement}. Hence, the bottom image is an SEM photo of the cross section of the contact between the laywood meta 5 material and the carbon black PLA.

The specimens presented in Fig. \ref{fig: Stacking Sequence}B were then attached to a black laser cut piece of Perspex with size of holes adjusted to the size and distribution of the specimens created to let as much light go through as possible when the actuators are curved. Fig. \ref{fig: Stacking Sequence}B shows the different positioning of the specimens evaluated. SVM aperture: Initial test with symmetric specimens. The top specimen is orientated downwards and the bottom specimen upwards. DSDF aperture: second test with both specimens orientated downwards. QSDF aperture: Initial geometries for Material Neural Network-like control of light. For the specimens presented in Fig. \ref{fig: Stacking Sequence}B SVM aperture and DSDF aperture, one hole 50 mm wide and 50 mm high at the centre of the black perspex plate is cut. For the specimens presented in Fig. \ref{fig: Stacking Sequence}B QSDF aperture, four holes of 40 mm wide and 25 mm high are cut, instead. For the specimens used in this last structure, the number of individual soft robotic material actuators was reduced from five (Fig. \ref{fig: Stacking Sequence}A to four. The specimens were fixed with an electrical clamp connected to a DC power supply and attached to the conditioning box via magnets.

\subsection{Validation of the Conditioning Setup}
Fig. \ref{fig: Stacking Sequence}C shows a schematic of the set-up developed to record the actuation of the specimens in Fig. \ref{fig: Stacking Sequence}A. The specimens have been inserted into a box made of perspex and completely opaque, apart from the transparent perspex face in front of the bowens\textregistered  LPD1 LED light panel and the Allied Vision Alvium 1800 U-234m camera. In the 55 mm-high box, a floor is positioned at 40 mm, beneath which two large Petri dishes filled with deionized water are placed. The humid air rises to the upper floor through square openings in the floor, where the light, humidity and temperature sensors are positioned. The temperature and relative humidity level inputs will be referred in \textcolor{inputgreen}{green} throughout this paper, outside of the introduction. The light measured with the sensor will be considered as the output of the test and is highlighted in \textcolor{outputorange}{orange} throughout the rest of this paper. Initial trials were carried out to verify the humidity inside the conditioning boxes. These trials confirmed that relative humidity stabilised in the conditioning boxes after 24 hours. As a result, all recordings began only after a minimum 24-hour waiting period following box closure.

The heating generated via the DC power supply shown in Fig. \ref{fig: Stacking Sequence}C is detailed in Fig. \ref{fig: Stacking Sequence}D and e. Thermal images have been obtained by positioning the specimens described in Fig. \ref{fig: Stacking Sequence}A in the box set up. A Flir A70 thermal camera is used to record images while the specimen is being heated. Images in Fig. \ref{fig: Stacking Sequence}D show the heat map of the specimen in Fig. \ref{fig: Stacking Sequence}E after 10 seconds and 13 minutes of heating with the DC power supply set up at 50.0 V on input. Fig. \ref{fig: Stacking Sequence}F shows the evolution of the maximum temperature on the surface of the specimen during heating. The temperature reaches a stable value after five minutes. The thermal images show constant values along the surface of the carbon black/PLA composite, showing no large defects present in the printed material. 

\newpage

\subsection{Design of the Material Actuating Neurone}
Since this was the first time the combination of material was 3D printed, the initial design required optimisation based on quantitative experimental measurements. For this reason, the performance of the two series of specimens in Fig \ref{fig: Stacking Sequence}B (SVM aperture and DSDF aperture) has been compared for benchmark purposes. The specimens have been positioned inside the testing box, then voltage inputs have been alternated between 0.0 V during at least 58 hours and 50.0 V for 10 hours. Light, temperature and humidity inside the box and images of the actuators have been detected using the experimental setup previously described in Fig \ref{fig: Stacking Sequence}C.

In the present context, reversibility is the capability of an actuation mechanism to obtain similar light changes between two actuator positions independently of the history of the actuator \cite{deKergariou2023TheChallenges}. This reversibility relates to the ability of the actuator to repeat actuation amplitude. Every reversibility data for hygromorph must be associated with an actuation history \cite{deKergariou2023TheChallenges}. This history is presented through a set of parameters, like the relative humidity variation n$_1$, the number of repeats n$_2$ and the duration of the conditioning n$_3$, the latter being defined as the difference between the actuation trigger start and the convergence value as presented in a previous publication \cite{deKergariou2023TheChallenges}. In the present study, the relative humidity is the main source of actuation, but the actuation trigger is the voltage. Therefore, the reversibility parameter n$_1$ is represented by the variation of the voltage. As a consequence, the parameters in the present study are n$_1$= 50 V, n$_2$=6 and n$_3$= 10 hours. To measure the effective reversibility of the structure, the light recorded by the light sensor serves as output. The term $\Theta_N$ in Eq. \ref{eq:degree_of_reversibility} provides the average of light output amplitude variation obtained over a given number of repeats.

\begin{equation}
    \Theta_N=\frac{\sum_{n=1}^{N} \delta Lx}{N}
    \label{eq:degree_of_reversibility}
\end{equation}

In Equation \eqref{eq:degree_of_reversibility}, $\delta Lx$ is the variation in light of the soft robotic material actuators between opened (with voltage off) and closed (with voltage on) configurations. \textit{N} is the number of iterations conducted during the reversibility test. The actuation speed of the material structure—here, the rate of light change—can be determined by two distinct properties \cite{deKergariou2023TheChallenges}: time to convergence, and time derivative of the output parameter. The time derivative of the output parameter is defined in the present study as the light variation, light varying speed, ($\xi_l$ in Equation \ref{eq:Lux_modulus}) immediately after turning on or off the voltage. The time to converge of the material actuating structure is defined as the time taken to reach 95\% of the final light output. For instance, in Fig. \ref{fig: Stacking Sequence}G this time corresponds to the first time when the light reaches inside the interval [0.95; 1.05]*$L_{1\_high}$ and is defined as $\iota_l$. For the decreasing curve this time corresponds to the interval [0.95; 1.05]*$L_{2\_low}$ and is referred to as $\iota_h$. 

\begin{equation}
    \xi_l=\frac{(l_{1\_low\_f}-L_{1\_low})}{(t_{1\_low\_f}-t_{1\_low\_i})}*100
    \label{eq:Lux_modulus}
\end{equation}

With $l_{1\_low\_f}$, $L_{1\_low}$, $t_{1\_low\_f}$ and $t_{1\_low\_i}$ parameters defined in Fig. \ref{fig: Stacking Sequence}G. This equation is also valid for any interval of time immediately after turning on the voltage with $l_{1\_high\_f}$, $L_{1\_high}$, $t_{1\_high\_f}$ and $t_{1\_high\_i}$.

An example of closing specimen is presented in Fig. \ref{fig: Stacking Sequence}H. The left image proposes a photo taken without voltage in the specimen. The middle image is a photo taken short time after voltage is switched on. This photo also displays an example of the tracker points used to obtain the width of one the ten actuators. Finally the right image presents a photo taken as the specimens are converging towards their closed position when the voltage is applied. The degree of closing is calculated from the images acquired during the experiment. The actuation strain is used to quantify the degree of closing of the specimens. Each acquired image is split between upper (top in Fig. \ref{fig: Stacking Sequence}H) and the lower (bottom in Fig. \ref{fig: Stacking Sequence}H) parts to separate the two specimens. Each part of the image is dissociated in five actuating parts treated separately. The image section of each actuator is assigned with tracking points using a Python script. To track points in space the OpenCV function \textit{goodFeaturesToTrack} was used with \textit{maxCorners} equal to 20000 and a \textit{blockSize} of 7. One of this actuating part and its tracking points are presented in Fig. \ref{fig: Stacking Sequence}H. Then these points had their position tracked during the experiment. The actuation width \textit{w} of the square containing all these point was measured along the experiment.

The actuation strain (representing the evolution of the actuation width \textit{w} defined in Fig. \ref{fig: Stacking Sequence}H is measured for each actuator with equation \ref{eq:strain}.

\begin{equation}
    \epsilon=\frac{(w_0-w_i)}{w_0}*100
    \label{eq:strain}
\end{equation}

In Eq. \ref{eq:strain}, $w_i$ represents the width of the actuator as displayed in Fig. \ref{fig: Stacking Sequence}H at a given time \textit{i}. The parameter d $w_0$ is the width measured on on the first image taken. The average is calculated from the five actuators of the specimen.

\subsection{Individual Behaviour of Material Actuating Neurone}
To assess the voltage-input/structure-response relationship, a separate experiment varied the voltage in the second actuator (from the top, Fig. \ref{fig: Stacking Sequence}B while keeping all other specimens unpowered. Fig. \ref{fig: Stacking Sequence}C also shows the measured output light and strain for each of the four actuators of the specimen considered in the QSDF aperture for this test. Equation \ref{eq:strain} is used to calculate the strain as a representation of the degree of opening \cite{deKergariou2025Hygromnemics:PreConstraining}.

\subsection{Control of the PAMMUNN via Data-Aware Back-Propagation Training}
\begin{figure}[h]
    \centering
    \includegraphics[width=0.983\textwidth]{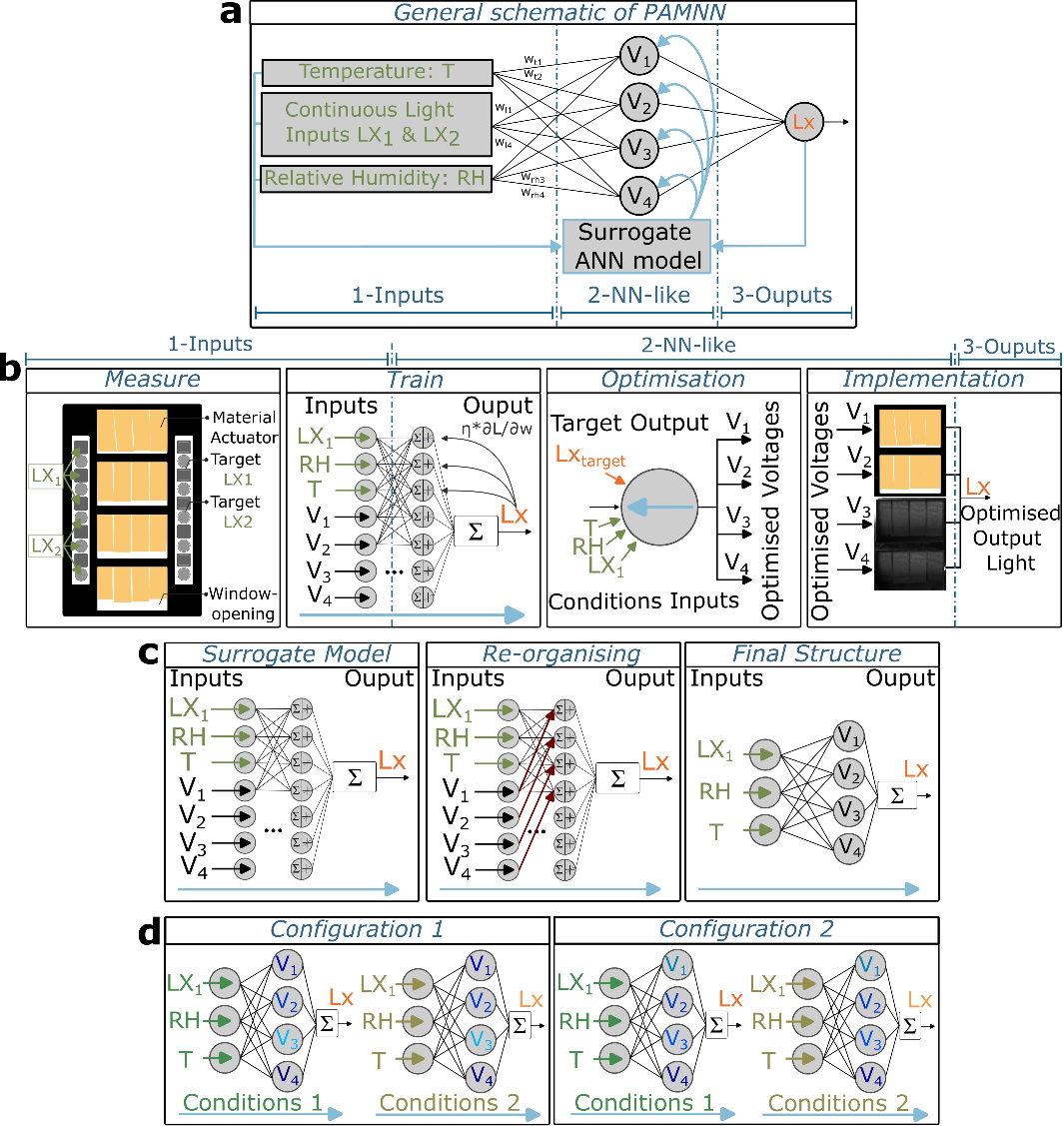}
    \vspace{-2.5mm}
    \caption{\textbf{Schematics of functioning of the artificial neural network-controlled structure via Data-Aware Back-Propagation Training and schematic of the creation of the physical-mode.} \textbf{A}, General schematic presenting the links between inputs (temperature, light outside, relative humidity) and the output (Light inside the box). \textbf{B}, Detailed schematics of the window-like structure and the calculations performed for its control. \textit{Measure:} view of the actuator from the camera, showing how the light targets used to measure LX$_1$ and LX$_2$ are positioned. \textit{Train:} Schematics for the training process of the neural network linking output and input. \textit{Optimisation:} schematics of the optimisation process to obtain voltage from the target light and the conditions of the system. \textit{Implementation:} Use of optimised voltages from the experimental set up to reach the desired output light.  \textbf{C}, Schematic of the optimiser re organisation to obtain PAMMUNN optimised configuration. \textbf{D}, Schematic of the two configurations selected to observe the behaviour of the PAMMUNN. }
    \label{fig:NN-like structure schematic}
\end{figure}

Fig. \ref{fig:NN-like structure schematic} presents the logic behind the control of the window-like structure via artificial neural network. The structure controlled is made of four actuating specimen placed one above the other, as the schematic of QSDF aperture presented in Fig. \ref{fig: Stacking Sequence}B presents. 

Fig. \ref{fig:NN-like structure schematic}A describes the target behaviour for the control of the window made with the physical adaptive material neural network. This figure illustrates that the physical nature of the inputs leads their quantities arriving on each specimen to be variable. For instance, heated air tends to rise as a result of density. Hence, $w_{t1}$, $w_{t2}$ are the weight coefficients representing this variation between the temperature received by each specimen and the general input temperature value considered for the control of the machine. Measuring exact temperature, humidity and light going to each actuating specimen is complex. Therefore, the approach implemented here is to measure a general value for each of the input parameter and observe the ability of the control strategy to account for these variations without precisely measuring them. The first step in implementation of the NN-like window structure is training (Fig. \ref{fig:NN-like structure schematic}B). During this step, a database is created by measuring the output light in the light sensor in Fig. \ref{fig: Stacking Sequence}C and linking it to the environmental parameters (input light, relative humidity and the temperature). The LX$_{1}$ and LX$_{2}$ light input parameters are considered, representing the RGB colour coefficients of gray targets positioned a few millimetres beside the NN-like window. This acquisition method was selected for its high repeatability and minimal impact on light transmission through the window. Fig. \ref{fig:NN-like structure schematic}B shows the position of the light targets used for measuring LX$_{1}$ and LX$_{2}$. The square shape target are used to measure LX$_{1}$ and are filled with dark grey (i.e. RGB values [98; 98; 98]). The circle shape targets are used to measure LX$_{2}$ and are filled with a light grey (i.e. RGB values [148; 148; 148]). The positioning of the camera for the acquisition of images is shown in Fig. \ref{fig: Stacking Sequence}C. The images are then uploaded in a Python script, as could be seen in Fig. \ref{fig:NN-like structure schematic}B a white circle is plotted in each circle target and a white square is drawn in each square light targets. The greyscale value of the pixel in the plotted circles and squares are then measured and averaged to obtain the value of LX$_1$ and LX$_2$, respectively. The database is used to train a single-layer neural network with seven nodes and a sigmoid activation function (Fig. \ref{fig:NN-like structure schematic}C). A gradient descent optimisation algorithm with a $\eta$=0.54 learning rate is used to adjust the weight of the NN layer. The loss function for the training is calculated as $L=\frac{1}{2}*(LX-LX_{target})^{2}$. In the training first the inputs and output were scale between 0 and 1. For the training, the LX$_1$ values were considered as input and the LX$_2$ were used to verify the quality of the LX$_1$ values obtained. As described in Fig. \ref{fig:NN-like structure schematic}B \textit{Sigmoid} functions are used for the layer of nodes.  The database obtained was divided in 80\% and 20\% for training and validation, respectively. To avoid overfitting data, the training was stopped after 2000 epochs if no improvement on the validation loss was achieved. In the training process described in Fig. \ref{fig:NN-like structure schematic}B, the voltages used to obtain the database and optimised range varied between 0.0 V and 35.0 V. As described in the Optimisation section of Fig. \ref{fig:NN-like structure schematic}B, the neural network structure was used as a surrogate model to predict output light. This approach described in Fig. \ref{fig:NN-like structure schematic}B uses the \textit{optimise} (method='\textit{L-BFGS-B}') function from the \textit{Scipy} package in Python is used to minimise the square of the difference between targeted light and the one obtained with a feed forward calculation using the weights determined during the NN training. The voltages derived are then implemented to achieve the desired light output specified by the user. This method is inspired by Physics-aware back propagation training used in PNN which requires a digital model to predict the behaviour of the system and optimised the controlling parameter \cite{Momeni2025TrainingNetworks}. Hence, we call data-aware back propagation training the method used here as it does not require a digital model but a database to predict the behaviour of the PAMMUNN machine. To limit the energy consumed and speed up the actuation, 30 optimised solutions have been identified for the voltages, and the solutions with the lowest sum between the four voltages was selected and implemented. In the Implementation section, the two bottom photos show the actuators closing as voltage is applied to their carbon black composite sections.

The database was acquired and its size was increased after each run by collecting data during the test presented in discussion. No less than 300 data points were used for the database used for the training of the neural networks presented in this study. Optimised voltage values were then obtained to reach desired output. The voltage output from the setup depends on both relative humidity and temperature. Therefore, real-time humidity and temperature readings cannot be used directly for optimisation, as they do not reflect the stabilised conditions achieved at the optimised operating point. To achieve optimal voltage output, we have to predict the stabilised temperature and humidity values. In order to achieve this, the correlation coefficients between the voltage values and the temperature and humidity were obtained with the Least Squares regression Python \textit{np.polyfit} function for the database of 300 data points (functions provided in supplementary materials). Initial voltage optimisation was then performed. Using the correlation coefficients, we then estimated the stabilised temperature and relative humidity at convergence. These updated values were reintroduced into the optimisation code for a second, refined voltage optimisation run. The values of voltage obtained were introduced again in the regression function to obtain the final predicted values of temperature and humidity to be used on the final prediction round with the trained Neural Network. Due to the limited correlation between light output and temperature the scale of the prediction was empirically limited. We adjusted the current temperature by adding only 10\% of the difference between the predicted and current values. When the optimised voltages were calculated, an overshoot was sent to the voltage to accelerate the convergence to the targeted interval. For each voltage, the difference between the current and the optimised values was calculated. To reach the target interval, 105\% (95\% for V4 in Fig. \ref{fig:NN-like structure schematic}B) of this difference was added to the optimised value for decreasing output light, and 180\% for increasing output light. Once the target interval was reached with the overshoot voltage values, the system applied the optimised voltage values to conclude the convergence of the output light.

\subsection{PAMMUNN: Incremental Learning}
For the control system shown in Fig. \ref{fig:NN-like structure schematic}, adaptability to its environment is crucial. van de Ven et al. define incremental learning as a system learning from a non-stationary stream of data \cite{vandeVen2022ThreeLearning}. The system described in Fig. \ref{fig:NN-like structure schematic} acquires new data by letting it evolve within its environment. For physical neural networks, database creation is time-dependent, which in turn affects the training process \cite{Momeni2025TrainingNetworks}. This test aims to demonstrate that the machine enhances its predictive accuracy over time by adapting its data input stream. In this study, learning involves exploring the design space to accommodate a wider range of conditions. To assess the system’s learning capability, the neural network’s predictive accuracy was evaluated based on the volume of acquired data. To achieve this, 35 points/configurations have been initially randomly selected from the database foe evaluation. The model has been then trained with increasing numbers of data points, starting from five then increasing by five until a database of 80 points. The interval of increase of the database has been then adjusted to 20 points for every evaluation, until the entire database was added for training out of the 35 points initially extracted for evaluation.

\subsection{PAMMUNN: Physical-mode}
In order to build a physical adaptive material neural network with less digitally dependent, the surrogate model should be removed from the control of the actuator during operation. This mode of operation relying less on digital apparatus it is named \textit{Physical-mode}. Fig. \ref{fig:NN-like structure schematic}C illustrates the configuration transition from the surrogate model-controlled structure discussed earlier to the physical-mode. Inspired by Lee et al. \cite{Lee2022MechanicalBehaviors}, who demonstrated mechanical neural network capabilities, these tests will identify two distinct configurations capable of producing identical outputs. Fig. \ref{fig:NN-like structure schematic}D presents a schematic of neural networks related to two configurations of specimen with voltages assigned. Each configuration is capable of obtaining the same outputs (one orange nuance for each output) when given similar set of inputs (one green nuance for each input). To achieve the required configurations, Lee et al. modelled the mechanical neural network and then run an optimisation on the model. To achieve Configurations 1 and 2 (Fig. \ref{fig:NN-like structure schematic}D), we applied the surrogate model optimization technique shown in Fig. \ref{fig:NN-like structure schematic}D. The neural network underwent identical training, with optimisation focused on minimising the sum of squared differences between the targeted and optimised light outputs under both conditions. The calculated voltage values were accepted only if the difference between individual voltages exceeded 3 V. For instance, in the example provided in Fig. \ref{fig:NN-like structure schematic}D the conditions targeted are: Conditions 1: (Lx: 180.3 Lx; LX$_1$: 185; T: 21.9 $\degree$C; RH: 91.0\%) Conditions 2: (Lx: 115 Lx; LX$_1$: 110.44; T: 22.3 $\degree$C; RH: 91.5\%). The optimisation allowed to obtain [0.23; 20.42; 0.01; 8.34] V and [23.32; 14.01; 6.06; 14.10] V for the two configuration making the prediction of light output (condition 1: 181.2 LX \& 114.0 LX) and (condition 2: 183.4 LX \& 110.1 LX), respectively. These testing parameter were the implemented in the QSDF set up presented in Fig. \ref{fig: Stacking Sequence} with the objective to validate the ability of the machine created to behave in the physical-mode.

\section{Acknowledgements}

C.d.K., A.W.P and F.S. acknowledge the support of the ERC-2020-AdG101020715 NEUROMETA project. The corresponding author, C.d.K.,also thanks the EPSRC Doctoral Prize Fellowship for supporting this work (Grant No. EP/W524414/1). 

\section{Data Availability Statement}
The data that support the ﬁndings of this study are available from the corresponding author upon reasonable request.
 
\medskip

%

\clearpage

\section{Supplementary Materials}

\subsection{Design of the Material Actuating Neurone.}
Fig. \ref{fig: Strain and light} shows the actuation strain, sensor light response, 95\% convergence time, light change speed, light variation amplitude, and reversibility for both up-down and DSDF aperture configurations. Due to the different geometric configurations between the up down and down down test setups, these two configurations are not compared against light intensity related parameters. However, the strain defined in Eq. \ref{eq:strain} is an indicator for the largest amplitude of opening in the down down test setup. Fig. \ref{fig: Strain and light}A and B show that for up down specimen the top row of actuators display an amplitude of strain not only larger than the bottom row of actuators but also of the row top row of actuator in the up down set up. This difference in actuation strain is due to the spatial distribution of the carbon black PLA and laywood material filaments. As shown in Fig. \ref{fig: Stacking Sequence}, in the DSDF aperture setup, the bottom specimen’s carbon black section is positioned near the laywood section of the top specimen. Thus, when current passes through the bottom specimen, its actuation—and that of the top specimen—is controlled via the heat radiation shown in Fig. \ref{fig: Stacking Sequence}D. Combining the effect of the carbon black material of the bottom and top specimen on the top specimen leads to large amplitude of actuation, consequently, large control potential for the output of the shading structure. Moreover, Fig. \ref{fig: Strain and light}B shows fewer instabilities compared to Fig. \ref{fig: Strain and light}A, where strain exhibits short-term fluctuations when the DC power supply is off. As a consequence, the neural network-like actuating machine has been made using only downwards-orientated specimens. This conclusion contradicts the design of self-shading surfaces proposed in previous studies, where specimen orientation was varied \cite{Reichert2015MeteorosensitiveResponsiveness,Battaglia2025DesigningPrinting}.

\renewcommand{\thefigure}{S\arabic{figure}}
\setcounter{figure}{0}
\begin{figure}[h]
    \centering
    \includegraphics[width=0.83\textwidth]{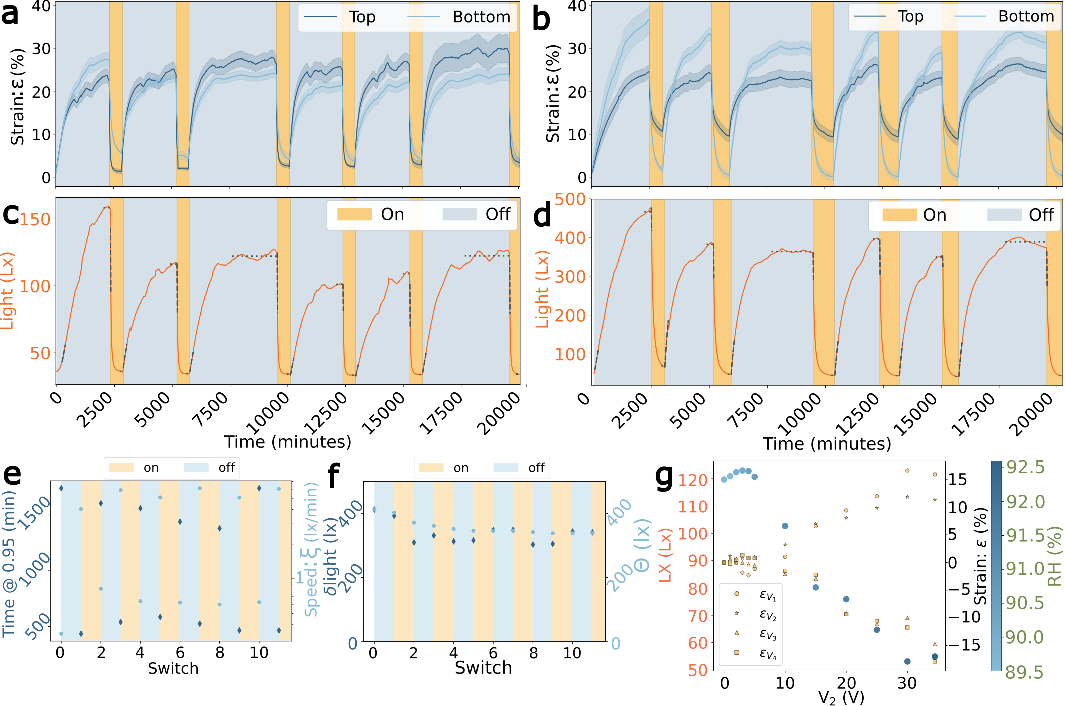}
    \vspace{-3mm}
    \caption{\textbf{Speed and repeatability of the SVM aperture and DSDF aperture specimens shown in Fig. \ref{fig: Stacking Sequence}B.} \textbf{A}, and \textbf{B}, The evolution of the actuation strain in the two rows of actuators (top and bottom) for the SVM aperture and the DSDF aperture, respectively. 
    \textbf{C}, and \textbf{D}, Evolution of the amount of light received by the light sensor for the SVM aperture and the DSDF aperture, respectively. "On": The electricity is running through the carbon black PLA filament, while the DC power supply is on. "Off": The DC power supply is off. The dark gray dashed and dotted lines represent the interval of data through which the speed ($\xi$) and time at 95\% data were measured, respectively. 
    \textbf{E}, Evolution of speed and time at 95\% convergence during reversibility test for the DSDF aperture. \textbf{F}, Evolution of the amplitude of light control and degree of reversibility during reversibility test for the DSDF aperture. \textbf{G}, Influence of the output light as function of the voltage input in the V$_2$ referenced specimen in Fig. \ref{fig:NN-like structure schematic}.}
    \label{fig: Strain and light}
\end{figure}

Fig. \ref{fig: Strain and light}D confirms that the stability of the actuation strain noted earlier for the DSDF aperture specimen is mirrored in the recorded light output. After the first 'Off' cycle (472 Lx), subsequent cycles record light levels within a 50 Lx range (355–405 Lx). This reduction stems from the decreased opening angle of the specimen, caused by humidity-induced changes of the material properties, such as the stiffness loss in natural fibre composites \cite{deKergariou2022TheComposites}. Cyclic conditioning alters the properties of natural fiber composites between the first and subsequent cycles. For example, Cadu et al. found that the stiffness of flax fibre-reinforced epoxy decreases, while the crystallinity of the flax fibres increases after the first conditioning cycle \cite{Cadu2019CyclicSensitive}. Fig. \ref{fig: Strain and light}E and F show the evolution of the two parameters characterising speed (light variation and time at 95\%), as well the reversibility for the opening and closing of the actuator in the DSDF aperture. Fig. \ref{fig: Strain and light}E shows that the speed is constant during drying after the first cycle of actuation. When the electricity is turned on, the speed of closing reaches values between 5 and 6 lx/min with standard deviation of 9.2\%. The 95\% of the final light output is obtained after an average of 8 hours and 22 minutes. However, the actuation rapidly gets close to a final value of 5.6 $\pm$ 0.47 lx/min  when the DC power supply is turned on, followed by a slower final convergence Fig. \ref{fig: Strain and light}B. It is critical to overshoot the target light level to accelerate actuation. The speed of opening of the actuators when the DC power supply is off 0.66 $\pm$ 0.08 lx/min. That speed is one order of magnitude lower than when the DC power supply is on and has similar repeatability with a coefficient of variation of 13.7\%. Fig. \ref{fig: Strain and light}F shows the evolution of the amplitude of the control light and the degree of reversibility of the structure. Fig. \ref{fig: Strain and light}F shows the presence of a reduced actuation amplitude after the first cycle, similarly to those shown in Fig. \ref{fig: Strain and light}D. Subsequent cycles exhibit stable actuation, with a 5.68\% coefficient of variation and an average light output of 325.8 $\pm$ 17.6 Lx. After the first cycle, the actuation amplitude stabilizes, causing the reversibility ($\Theta$) to decline until it plateaus at the fifth switch. This indicates consistent opening/closing behaviour in subsequent cycles.

\subsection{Individual Behaviour of Material Actuating Neurone}
Fig. \ref{fig: Strain and light}G illustrates how the output light varies with voltage changes in the specimen (V$_2$), as referenced in Fig. \ref{fig: Strain and light}G. Fig. \ref{fig: Strain and light}G demonstrates that increasing V$_2$ from 0 V to 3 V raises output light by 3.3 Lx, while higher voltages (up to 30 V) reduce it to 53.1 Lx. This confirms that higher voltages dry the actuator, causing it to close further and transmit less light. In the QSDF aperture, the two top specimens—especially the one nearest V$_2$—close with increasing voltage (positive strain), while the two furthest from V$_2$ open (negative strain) due to their lack of exposure to local drying effects. However, as shown in Fig. \ref{fig: Strain and light}G, increasing voltage raises the relative humidity (RH) inside the box, triggering standard hygromorphic behaviour \cite{deKergariou2025Hygromnemics:PreConstraining}, hence the V$_3$ and V$_4$ opening of the the actuators (Fig. \ref{fig:NN-like structure schematic}).

\subsection{Interpolation function between the voltages and the relative humidity and temperature}
RH and T represent the relative humidity and temperature, respectively. V$_1$, V$_2$, V$_3$ and V$_4$ are the four voltages described throughout the paper. The fitted functions are presented in the following equations.

RH = 0.015*$V_1$ + 0.046*V$_2$ + 0.033*V$_3$ + 0.051*V$_4$ + 89.6

T = 0.014*V$_1$ + 0.021*V$_2$ + 0.034*V$_3$ + 0.051*V$_4$ + 20.78

\subsection{Extra example of Scenario 1 from 6.3.1}

\begin{figure}[h]
    \centering
    \includegraphics[width=0.983\textwidth]{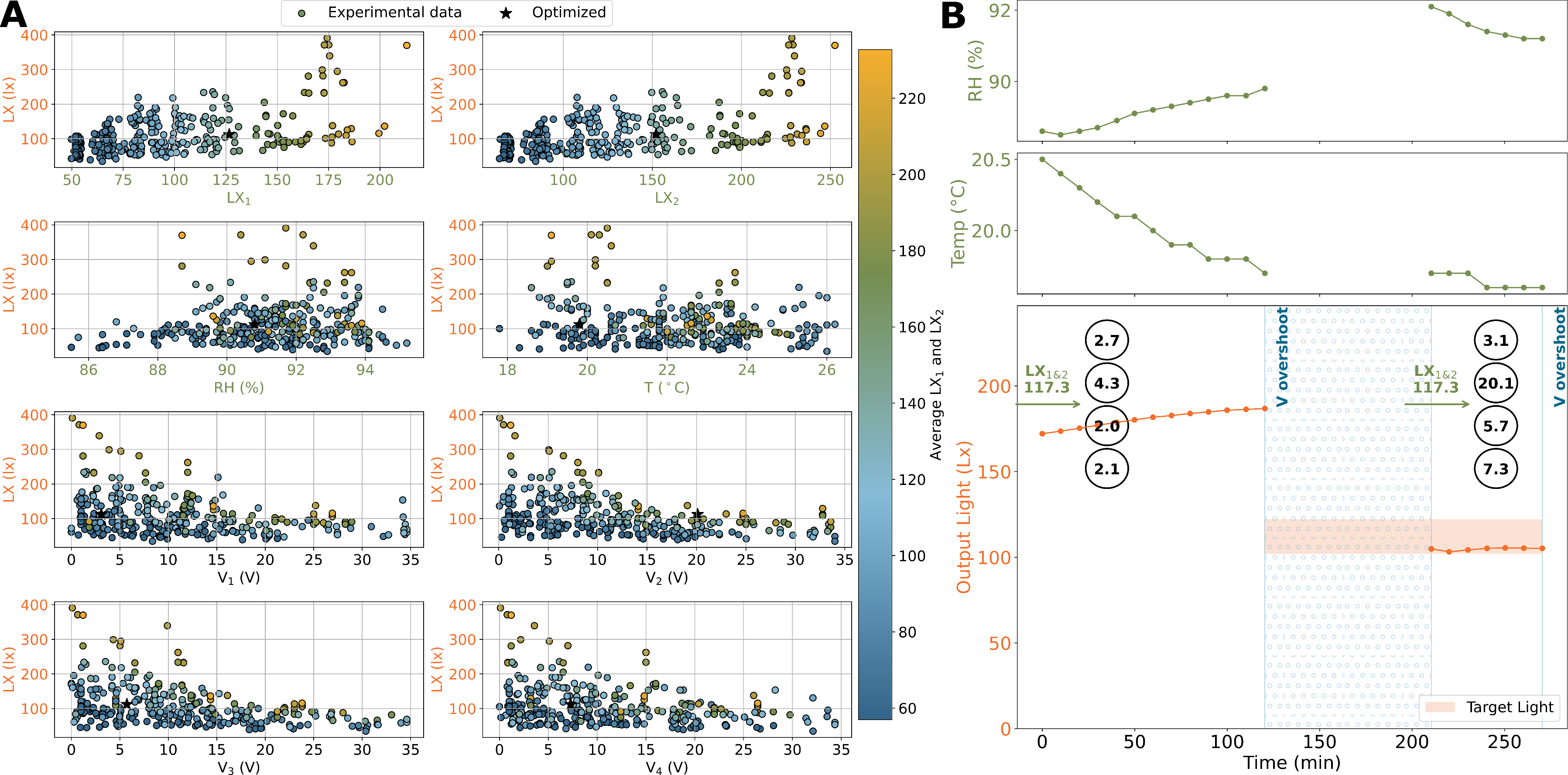}
    \vspace{-3mm}
    \caption{PAMMUNN via Data-Aware Back-Propagation Training: Scenario 1 - Example 3. The prediction conducted to obtain the output presented in \textbf{a} was conducted in Fig. \ref{fig: database example}.}
    \label{fig: PAMMUNN via Physics-Aware Back-Propagation Training: Scenario 1 - Example 3}
\end{figure}

\begin{figure}[h]
    \centering
    \includegraphics[width=0.983\textwidth]{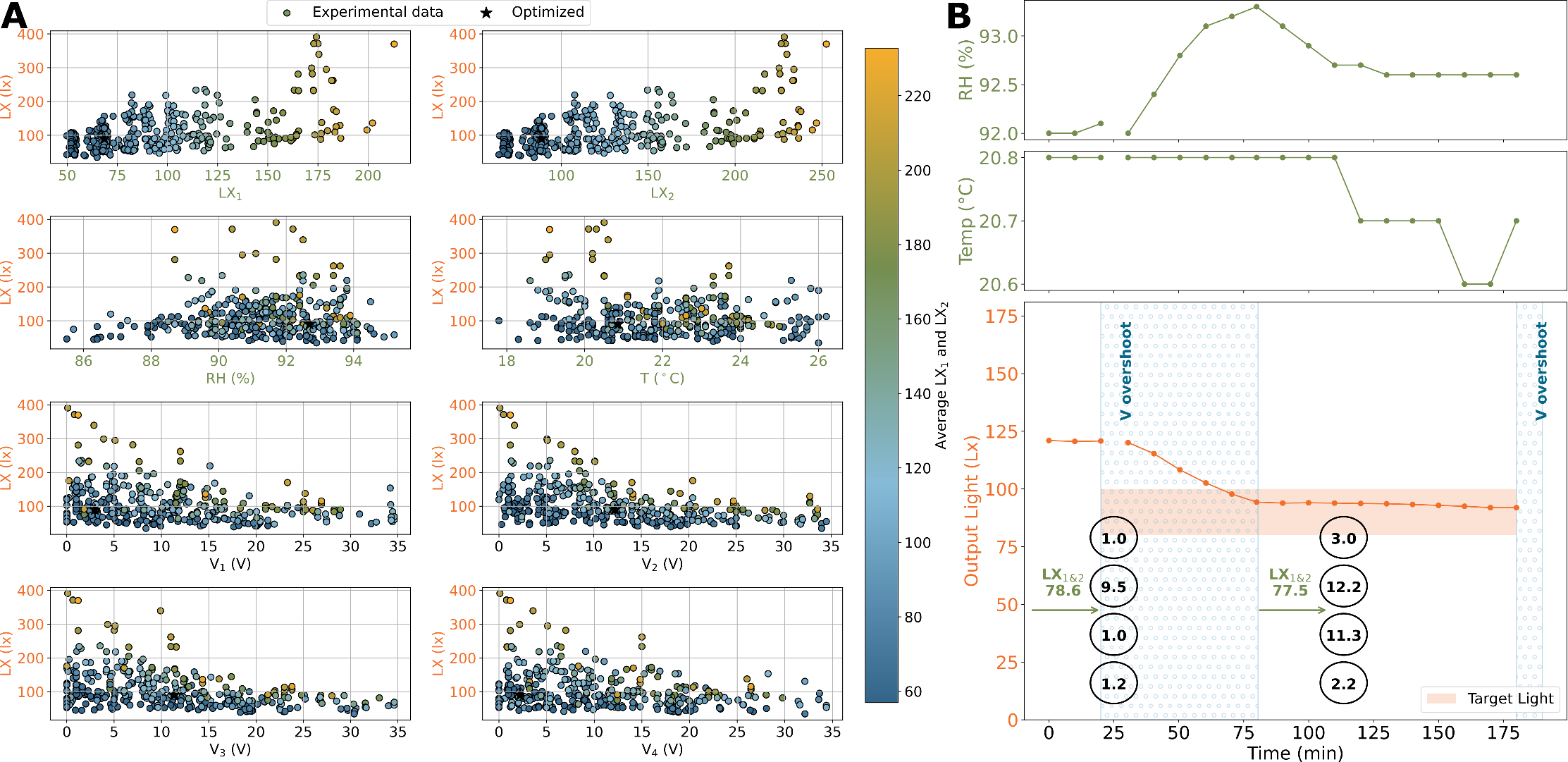}
    \vspace{-3mm}
    \caption{PAMMUNN via Data-Aware Back-Propagation Training: Scenario 1 - Example 4}
    \label{fig: PAMMUNN via Physics-Aware Back-Propagation Training: Scenario 1 - Example 4}
\end{figure}

\begin{figure}[h]
    \centering
    \includegraphics[width=0.983\textwidth]{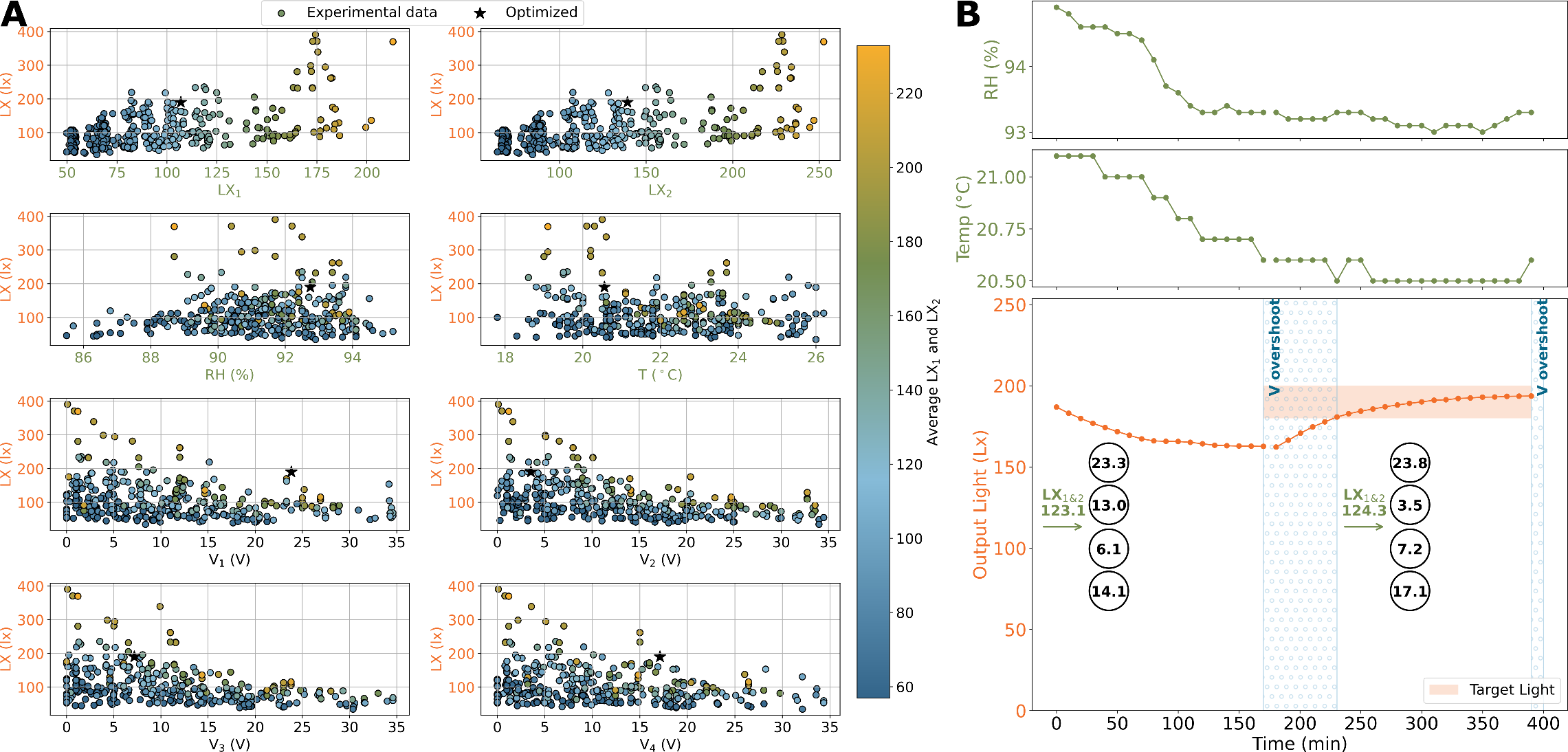}
    \vspace{-3mm}
    \caption{PAMMUNN via Data-Aware Back-Propagation Training: Scenario 1 - Example 5}
    \label{fig: PAMMUNN via Physics-Aware Back-Propagation Training: Scenario 1 - Example 5}
\end{figure}

\clearpage

\subsection{Extra example of Scenario 2 from 6.3.2}
\begin{figure}[h]
    \centering
    \includegraphics[width=0.983\textwidth]{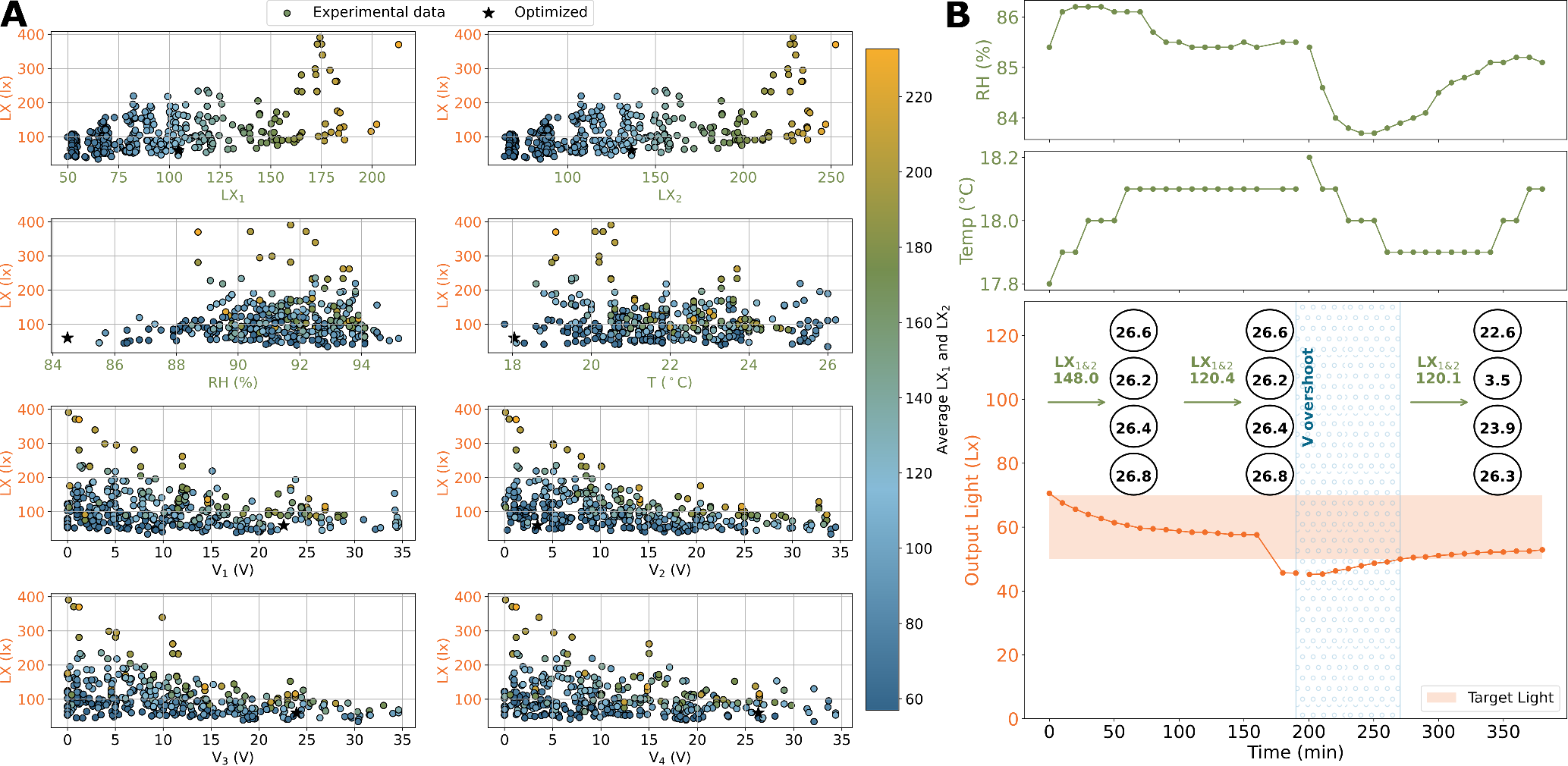}
    \vspace{-3mm}
    \caption{PAMMUNN via Data-Aware Back-Propagation Training: Scenario 2 - Example 2}
    \label{fig: PAMMUNN via Physics-Aware Back-Propagation Training: Scenario 2 - Example 2}
\end{figure}

\begin{figure}[h]
    \centering
    \includegraphics[width=0.983\textwidth]{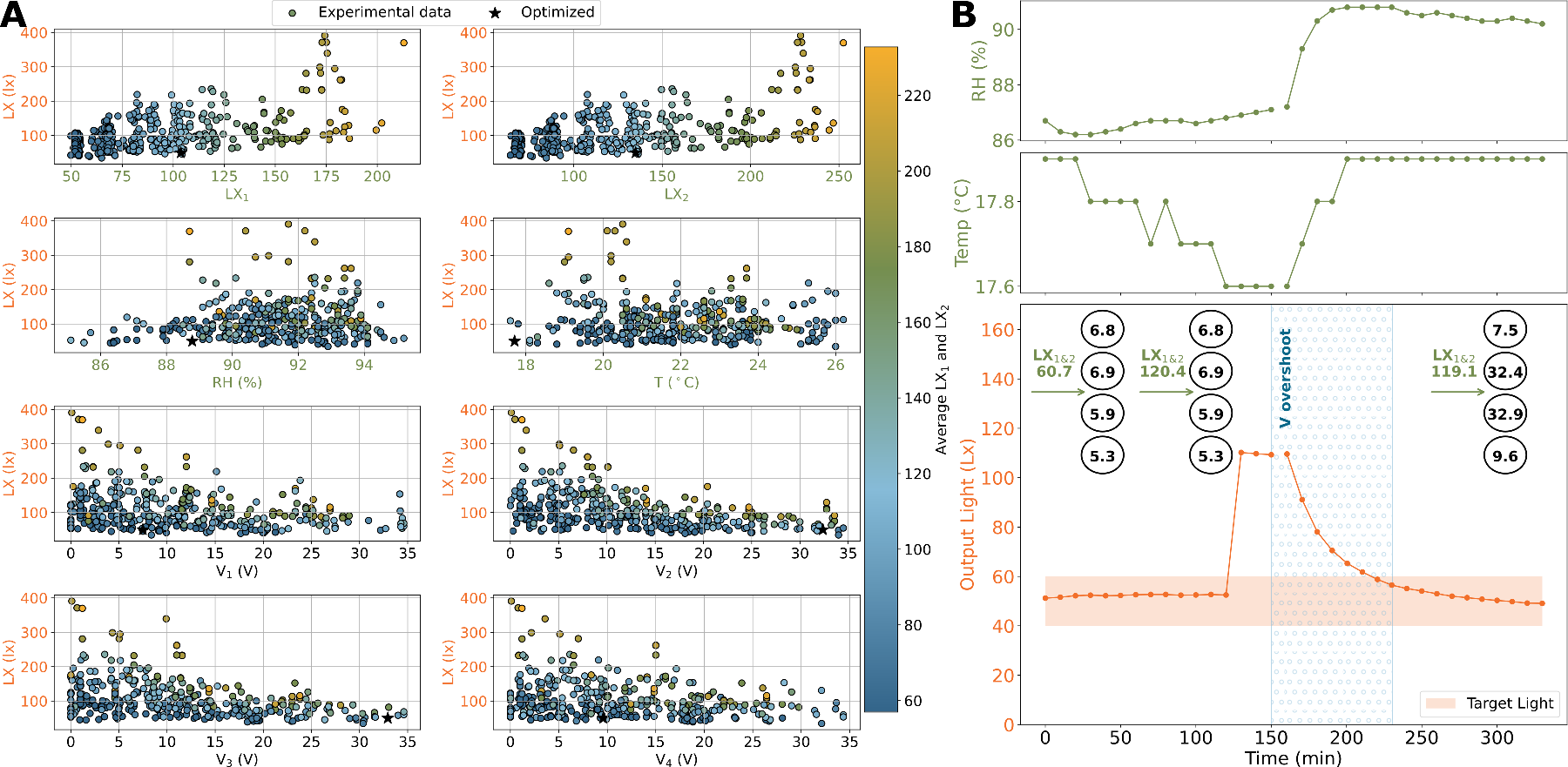}
    \vspace{-3mm}
    \caption{PAMMUNN via Data-Aware Back-Propagation Training: Scenario 2 - Example 3}
    \label{fig: PAMMUNN via Physics-Aware Back-Propagation Training: Scenario 2 - Example 3}
\end{figure}

\begin{figure}[h]
    \centering
    \includegraphics[width=0.983\textwidth]{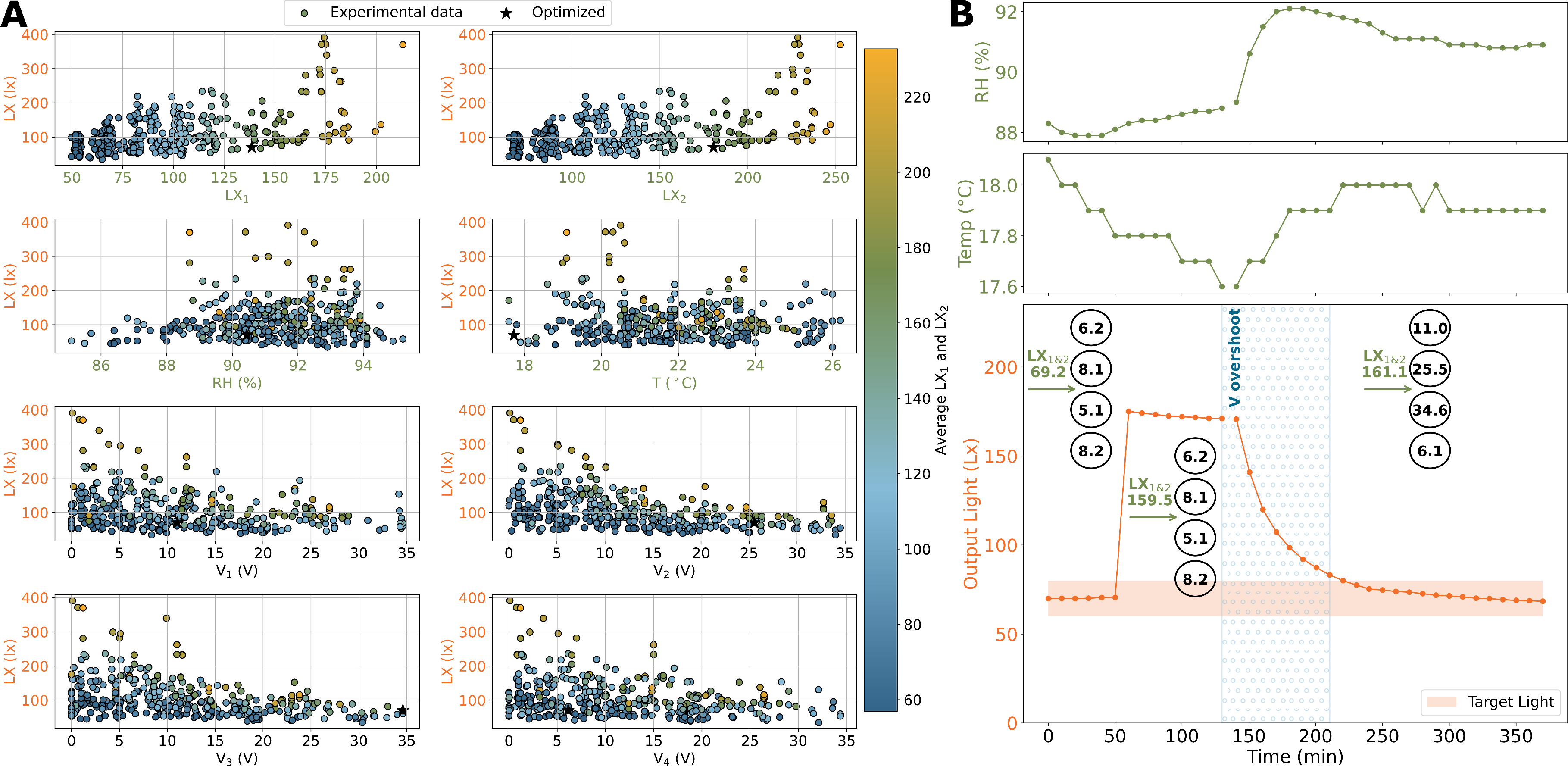}
    \vspace{-3mm}
    \caption{PAMMUNN via Data-Aware Back-Propagation Training: Scenario 2 - Example 4}
    \label{fig: PAMMUNN via Physics-Aware Back-Propagation Training: Scenario 2 - Example 4}
\end{figure}

\clearpage

\subsection{Extra example of Increase Autonomy of 6.5.}
\begin{figure}[h]
    \centering
    \includegraphics[width=0.983\textwidth]{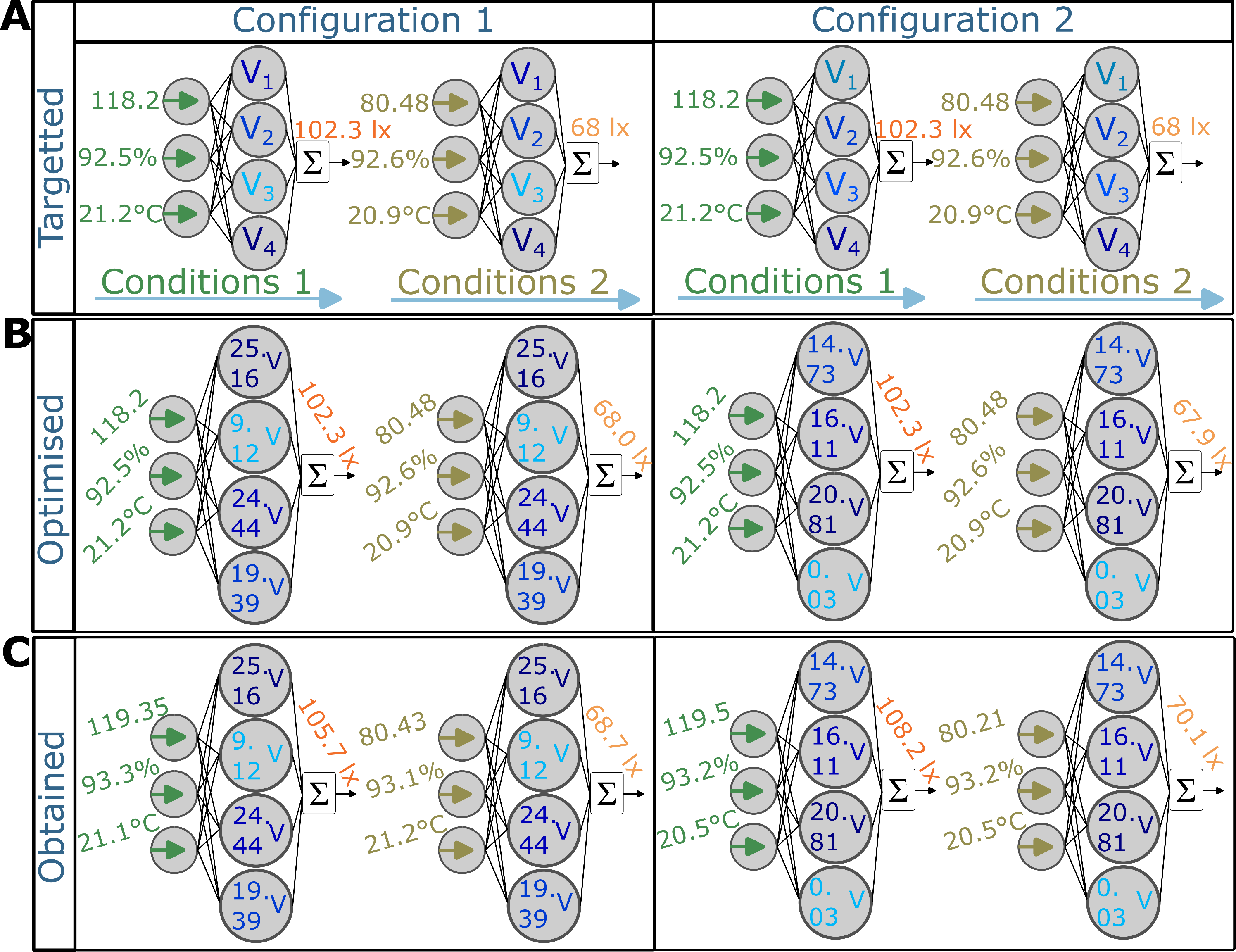}
    \vspace{-3mm}
    \caption{Physical-mode for PAMMUNN: Example 2}
    \label{fig: Optimised Configurations for PAMMUNN: Example 2}
\end{figure}

\begin{figure}[h]
    \centering
    \includegraphics[width=0.983\textwidth]{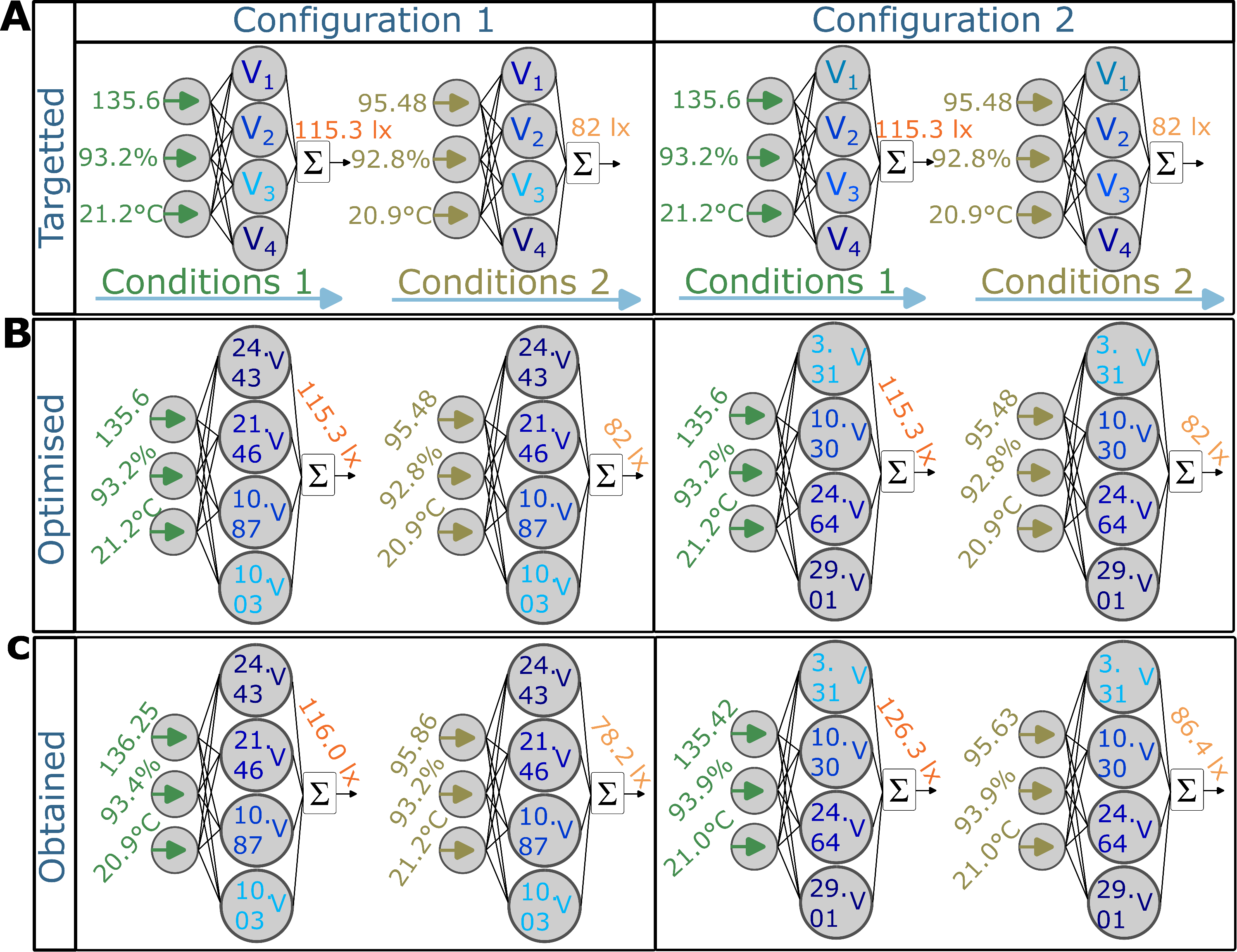}
    \vspace{-3mm}
    \caption{Physical-mode for PAMMUNN: Example 3}
    \label{fig: Optimised Configurations for PAMMUNN: Example 3}
\end{figure}

\end{document}